\documentclass[12pt,preprint]{aastex}

\newcommand{\vsxf}{V2492~Cygni}
\newcommand{\vsx}{V2492~Cyg}

\slugcomment{Version: \today}

\shorttitle{Gemini Observations of \vsxf}
\shortauthors{Aspin \& Reipurth}

\begin{document}

\title{V2492~CYGNI: THE EARLY EVOLUTION OF THE 2010 OUTBURST}

\author{
Colin~Aspin\altaffilmark{1}}

\altaffiltext{1}{Institute for Astronomy, University of Hawaii at
  Manoa, 640 N. A'ohoku Place, Hilo, HI 96720. {\it
    caa@ifa.hawaii.edu, reipurth@ifa.hawaii.edu}}

\begin{abstract} 
  We present Gemini-North optical and near-IR observations of a young
  eruptive star in Cygnus, designated as \vsxf\ in the General Catalog
  of Variable Stars.  This object is one of two young stars, located
  within 2 degrees of each other, that recently brightened by around
  5~mags and were reported as possible new FU~Orionis-type variables.

  The outburst spectrum of \vsxf\ shows atomic emission features
  throughout the optical and near-infrared.  In the optical, H$\alpha$
  is in emission and has an associated blue-shifted absorption
  component.  The far-red \ion{Ca}{2} triplet lines are also in
  emission and, as with H$\alpha$, possess significant blue-shifted
  absorption.  The optical TiO molecular bands are also in emission.
  In the near-infrared, Pa$\beta$, Br$\gamma$, and the CO overtone
  bandheads are strongly in emission.  Such spectral characteristics
  are very similar to those exhibited by both EX~Lupi, the progenitor
  of the EXor class of eruptive variables, during its 2008 extreme
  outburst, and V1647~Ori during its elevated phase in 2003.

  Additionally, we consider archival data on \vsxf\ and investigate
  the pre-outburst nature of this young star.  We construct a
  quiescent-phase spectral energy distribution which, via model
  fitting, gives insight into the circumstellar environment of the
  object prior to the current eruption.

  Our investigation strongly suggests that the \vsxf\ outburst appears
  to be more consistent with an EXor event, a fact supported by its
  recent 2~mag fading and stochastic variability.

\end{abstract}


\section{INTRODUCTION}
During the month of August 2010, eruptive outbursts of the two young
stars HBC~722 and \vsxf\ were reported by Semkov \& Peneva (2010a,b)
and Itagaki (2010), respectively.  Both stars are located in the
extensive Cygnus star formation complex near the North America and
Pelican nebulae (see Reipurth \& Schneider 2008 and references
therein), and both were found to have optically brightened by between
4 and 6 magnitudes. In the case of HBC~722, this brightening occurred
over a period of around three months, while for \vsx, the change in
brightness occurred over 10 months.  \vsx\ was independently
discovered by the Palomar Transient Factor (PTF) and labeled PTF10nvs
in a detailed study by Covey et al. (2011).  The American Association
of Variable Star Observers, the AAVSO, designate the star
VSX~J205126.1+440523.

These two independent events are significant because such large
amplitude outbursts are rare occurrences.  Two types of eruptive
events are known in young stellar objects (YSOs), specifically, those
named after FU~Orionis (Wachmann 1954; Herbig 1966) and those named
after EX~Lupi (Herbig 1977).  FU~Orionis outbursts, often referred to
as FUor outbursts (Ambartsumian 1971), show 5--6 magnitude increases
in optical brightness and can take from decades to even a century to
fade back into quiescence.  The two most most recent FUor events
observed were the outbursts of V1057~Cyg in 1969 (Welin 1971; Herbig
\& Harlan 1971) and the eruption of V1647~Ori in 2004 (e.g., Reipurth
\& Aspin 2004, Briceno et al. 2004, Aspin \& Reipurth 2009).  The
latter has had an unusual light curve, fading away after only two
years, but returning again to its elevated state, and today the
available evidence supports its classification as a FUor.  EX~Lupi
outbursts, commonly called EXors (Herbig 1977, 1989), exhibit optical
brightness increases of 2--5 mags and fade after periods of a few
months to a few years (Herbig 1977, 2008).  They are also observed to
be repetitive.  For example, the EXor variable VY~Tau brightened and
faded more than a dozen times between 1930 and 1970 (Meinunger 1969,
1971; Herbig 1977, 1990).

Both types of eruptive events are thought to be driven by the same
underlying process, commonly believed to be enhanced mass accretion
through a disturbed/disrupted circumstellar disk and onto the young
stellar photosphere.  Quiescent T~Tauri star accretion rates of
10$^{-7}$ to 10$^{-8}$~M$_{\odot}$~yr$^{-1}$ have been observed to
increase by a factor of 100 to 1000 during eruptive outbursts and can
thereby result in a significant amount of mass being accreted over
relatively short timescales.  This so-called ``accretion disk model''
(Hartmann \& Kenyon 1985, 1996) can explain many of the observed
properties of both FUors and EXors, although the mechanism by which
the disk becomes unstable is still being debated.  An outstanding
question regarding FUor and EXor outbursts is why they differ so much
in many of their observed characteristics including, for example, the
event timescale and amplitude, and their optical and near-infrared
(NIR) spectral properties (FUors exhibit predominantly optical and NIR
absorption spectra whereas EXors always possess strong emission
features).  An alternative explanation for FUor outbursts has been
proposed by Herbig (1989), Herbig et al. (2003), and Petrov \& Herbig
(1992, 2008).  Their model, which is based on a rapidly rotating star
with regions of significantly different photospheric temperatures, can
also explain many of the observed characteristics of FUor eruptions.

In this paper we present optical and NIR observations of the recently
discovered eruptive variable \vsx. The star is located in a small
molecular cloud core surrounded by a bright rim in the Lynds Bright
Nebula LBN~343.  Our goals are to {\it i)} characterize the early
stages of the eruption, {\it ii)} better understand how such a young
star fits into the established framework of FUor and EXor outbursts,
{\it iii)} provide new observations that can be used to further test
theoretical/numerical models, and {\it iv)} gain further insight into
the initiation and progression of such dramatic and highly energetic
events.

\section{OBSERVATIONS AND DATA REDUCTION}
Table~\ref{obslog} shows the complete observing log for the
observations presented below.  All observations of \vsx\ were taken
using the ``Fredrick C.  Gillett'' Gemini-North 8~m telescope located
on Mauna Kea, HI, in September, October, and November 2010 (see
Table~\ref{obslog} for complete dates).  The instrument used for the
optical observations was GMOS-N, the facility optical imager and
spectrograph (Davies et al.  1997; Hook et al. 2004).  For the
imaging, we used the r' filter and a total exposure time per source of
60~s.  For the spectroscopy, we used the B600 grating and central
wavelength 7500~\AA.  A 0$\farcs$75 wide long-slit was used resulting
in a resolving power, {\it R}, of $\sim$1200 (0.45~\AA~pixel$^{-1}$).
This value of {\it R} gives a full-width half maximum (FWHM) of
unresolved lines of $\sim$130~km~s$^{-1}$.  The total on-source
exposure time for the spectroscopy was 300~s.  Identical observations
of the spectrophotometric standard star G191-B2B were also taken to
allow flux calibration and sensitivity function definition.  All
spectra were reduced using the Gemini GMOS {\it iraf} package (v1.10).

Our NIR imaging observations were taken over the aforementioned period
using the Gemini-North facility NIR camera, NIRI (Hodapp et al. 2003).
Due to the intrinsic brightness of the sources, the f/32 camera was
used (allowing shorter exposure times) and data was acquired in J, H,
K', and L' filters.  For each filter, four dithered images were taken
to allow for both sky subtraction and uncertainty estimation.  Total
exposure times of 20~s were used for J, H, and K', and 48~s for L'.
Similar observations of the NIR faint standard star FS~150 were also
taken for flux calibration purposes.  The images were reduced using
the Gemini NIRI {\it iraf} package (v1.10).  Extraction of aperture
photometry from the reduced images was performed using the Starlink
{\it gaia} program.

NIR spectroscopic observations were acquired using the Gemini-North
facility integral-field unit (IFU) spectrograph, NIFS (McGregor et al.
2003).  Observations using the J, H, and K gratings were taken with
total exposure times of 120~s per waveband and resulted in spectra
with a spectral resolution of {\it R}$\sim$5000.  Sky observations,
taken using an 'ABBA' offset sequence, were also acquired to
facilitate accurate sky subtraction.  Similar observations of the A0~V
star HIP~103222 were taken to allow the removal of telluric features
from the target spectra.  The data were reduced using the Gemini NIFS
{\it iraf} package (v1.10).  The final spectra of the targets are the
sum of the IFU pixels lying within an 0$\farcs$5 radius software
aperture centered on the targets.  Flux calibration was performed
using the broad-band J, H, and K source brightness close to the time
of the spectroscopic observations.  We converted the broad-band
magnitudes to flux density using the Gemini Observatory conversion
tool\footnotemark
\footnotetext{http://www.gemini.edu/sciops/instruments/midir-resources/imaging-calibrations/fluxmagnitude-conversion}.

\section{\vsx\ AND ITS ENVIRONMENT}
Below we consider our knowledge of \vsx\ prior to its 2010 outburst.  
In what follows, we use the phrase {\it 'post-outburst'} to refers to
the period from immediately after the rise to maximum light but prior
to the sources return to its pre-outburst, quiescent brightness.  

\subsection{At Optical Wavelengths}
\begin{figure}[t] 
\begin{center}
\includegraphics*[angle=0,scale=0.8]{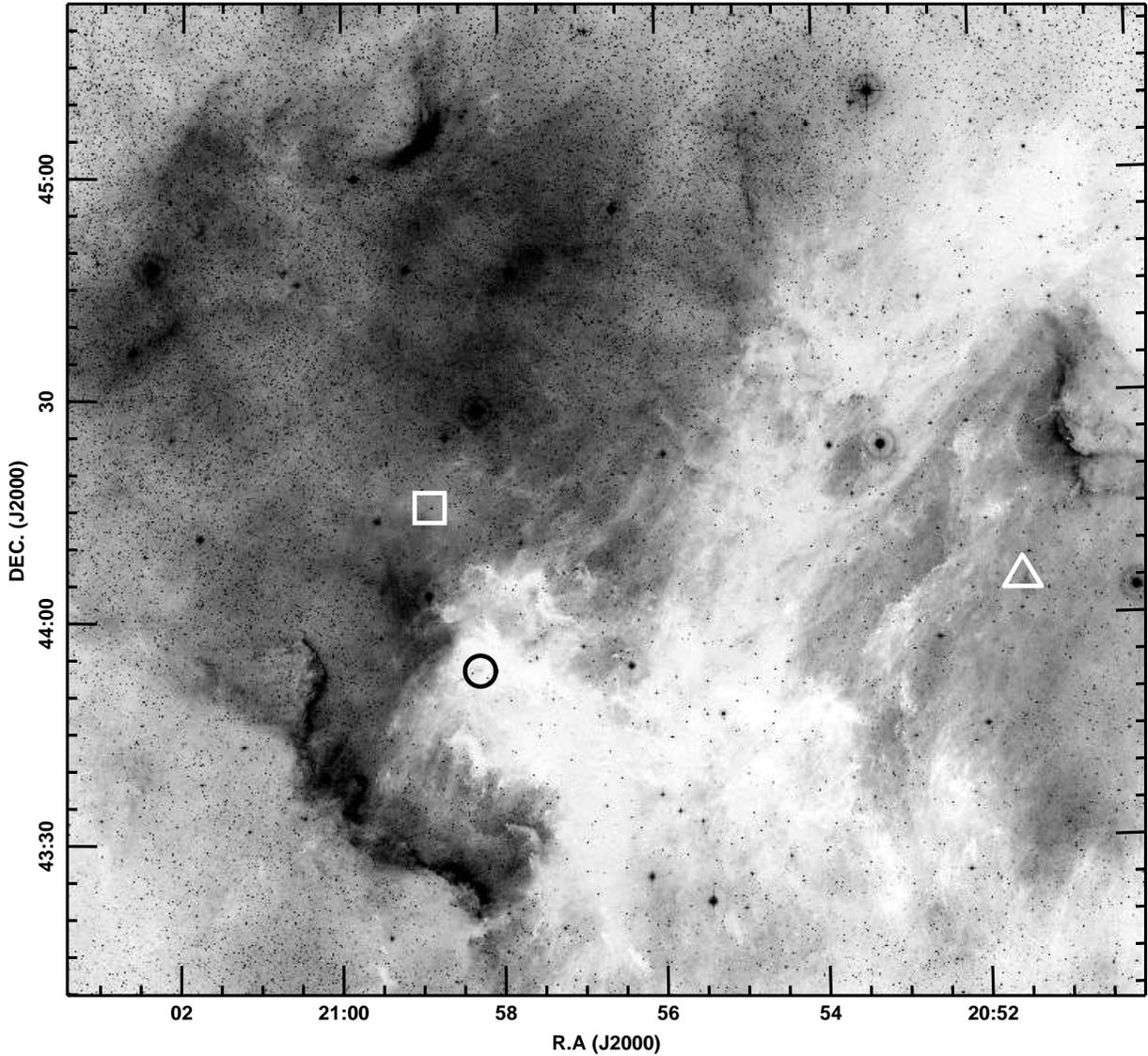} 
\caption{The DSS-2 red plate image of the region surrounding 
\vsx.   The image is approximately 2 degrees square.  The color table
of the image has been reversed so that bright objects are black and dark 
object white and hence the dark clouds associated with this region are white.
The location of \vsx\ is indicated by the white triangle.  
The white square indicates the classical FU~Orionis eruptive variable 
V1057~Cyg while the black circle shows the second young star in the region 
that recently erupted, HBC~722. The bright complex (black) to 
the north is the North America nebula (NGC~7000) while the one to the 
west is the Pelican Nebula (IC~5070/5067).
\label{dssr}}

\end{center}
\end{figure}

\begin{figure}[t] 
\begin{center}
\includegraphics*[angle=270,scale=0.8]{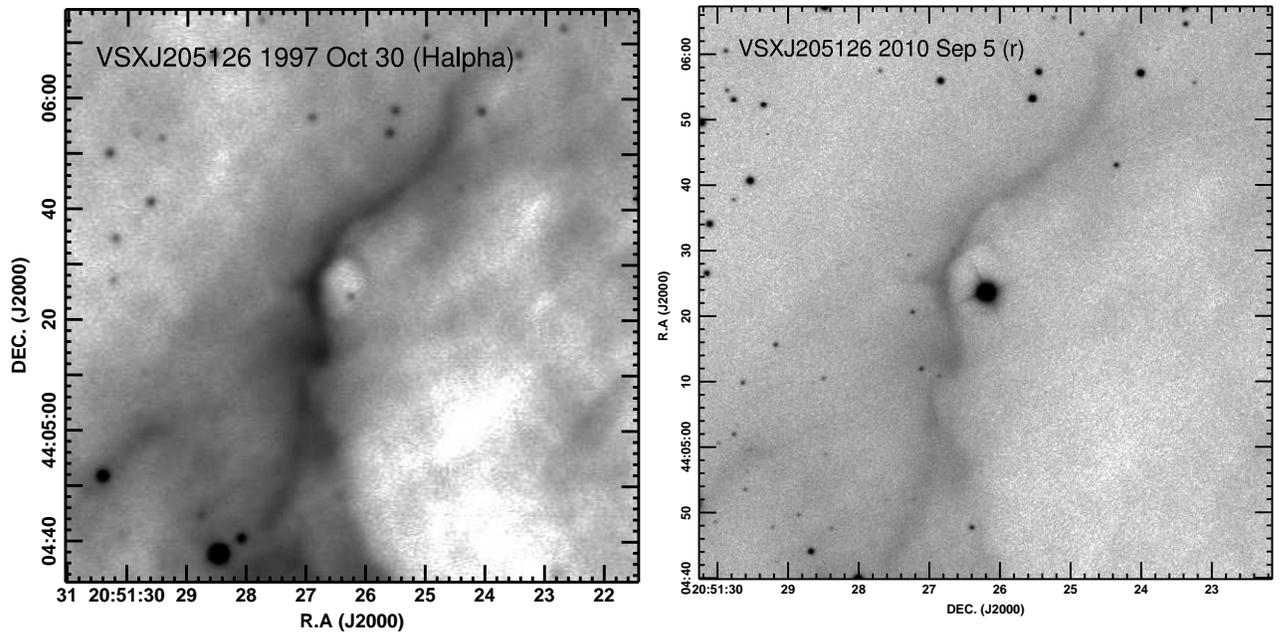} 
\caption{Optical images of \vsx.  The left and right images are
of \vsx\ (in the4 center of the image) from UT 1997 Oct 30 and 
UT 2010 Sep 5, respectively.  
The 1997 image is from Bally \& Reipurth (2003) and is an
H$\alpha$ image.  The 2010 image is in r' and from Gemini-North.  Note the 
significant brightness difference of \vsx. 
\label{ims}}

\end{center}
\end{figure}

In optical archival images taken in October 1997, \vsx\ appears as a
faint, unremarkable star located close to the edge of an irradiated
molecular cloud edge around 15$'$ southeast of the Pelican Nebula (see
Figure~\ref{dssr}).  It is associated with an {\it IRAS} source,
IRAS~20496+4354, and is the possible exciting source of a Herbig-Haro
bow shock, HH~569, some 2$'$ (0.35~pc) to the south (Bally \& Reipurth
2003).  Images of the region immediately surrounding \vsx\ are shown
in Figure~\ref{ims}.  Bally \& Reipurth (2003) classified
IRAS~20496+4354 as a moderate-luminosity Class~I protostellar object.
In the Dobashi et al. (2005) visual extinction (A$_V$) maps, \vsx\
lies near a filamentary dust structure in a region with a general
line-of-sight A$_V$ of $\sim$4--5~mags.  The distance to the North
America/Pelican Nebula region of Cygnus has been estimated to be
550$\pm$50~pc (Laugalys et al.  2006).

Itagaki (2010) recently documented eruptive behavior in \vsx\ finding
that its optical (unfiltered) brightness rose from $>$17.5 on
UT~2008~Aug~7 to 13.8 two years later.  The commencement of this
outburst, however, could have been any time between UT~2008~Aug~7 and
his next observation on UT~2009~Dec~19.  His multiple observations
between 2009.9 and 2010.7 show a close to linear rise in optical
brightness of approximately 0.25~mag~month$^{-1}$.  For reference,
USNO-B1 gave optical magnitudes for \vsx\ of B=20.17 and R=18.28 on
1979.9.  Covey et al. (2011) presented an extensive, detailed study of
\vsx\ including considerable R-band optical photometry from the
Palomar Transient Factory images.  These data demonstrated that the
current event peaked in August 2010 (at R=13.3) and rather than the
linear rise in brightness as seen by Itagaki (2010), significant
stochastic variability was present from as early as August 2009.  In
addition, their light curve showed multiple brightness peaks with a
typical peak-to-peak timescale of $\sim$100~days suggesting the
possibility of periodic as well as uncorrelated variability.

\subsection{At Infrared Wavelengths}
\begin{figure}[t] 
\begin{center}
\includegraphics*[angle=270,scale=0.6]{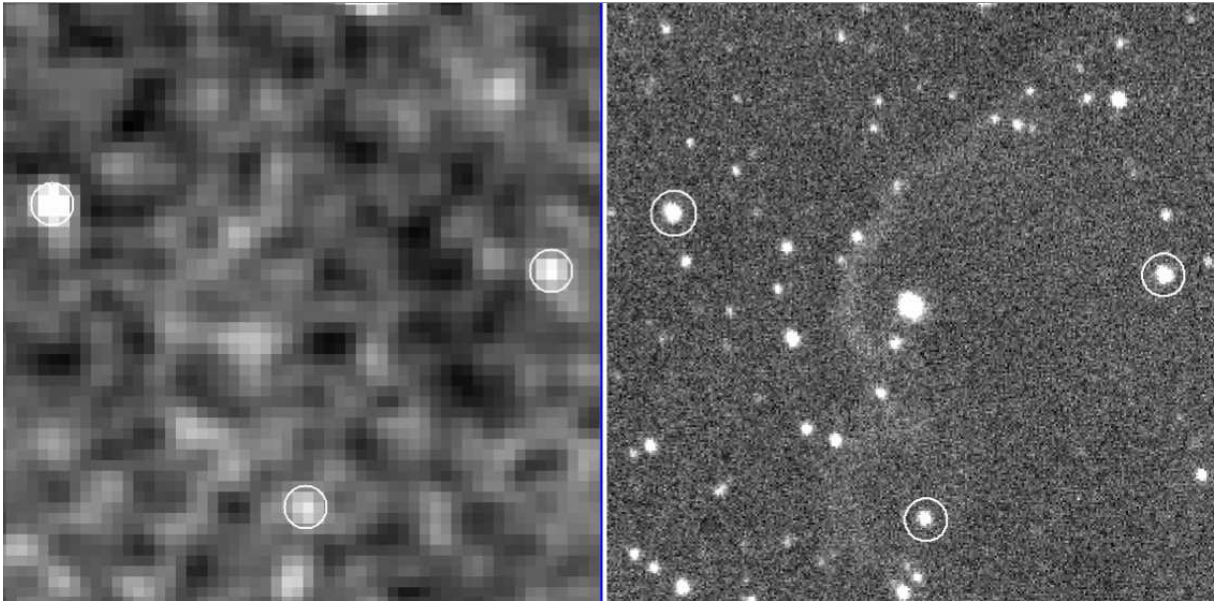} 
\caption{Comparison of the {\it 2MASS} K' image (left, taken in 2000) 
of the region containing \vsx\ (left) and the UKIRT UKIDSS K' 
image (right, taken in 2006). North is at the top, east to the left and 
each image is approximately 60$''$ square. Common stars are 
indicated by white circles.  Note that there is no sign of \vsx\ in 
the {\it 2MASS} images indicating that it is fainter than the 
10$\sigma$ K' detection limit of 14.3~mags.  For reference, the western 
of the circled stars has a K' magnitude  of 14.7~mags.
In the UKIDSS Galactic Plane Survey, \vsx\ has a K' magnitude of 13.15.
\label{2mass-ukidss}}

\end{center}
\end{figure}

\vsx\ was not detected in the {\it 2MASS} survey at any wavelength.
This implies that at the time the survey images were acquired, the
source was considerably fainter than the quoted {\it 2MASS} 10$\sigma$
detection limits of 15.8, 15.1, and 14.3~mags at J, H, and K',
respectively (Skrutskie et al.
2006)\footnotemark\footnotetext{Figure~15 of Skrutskie et al (2006)
  indicates that the 3$\sigma$ detection limit of {\it 2MASS} is
  approximately 0.5, 1.0, and 1.4~mags fainter at J, H, and K',
  respectively.  In the {\it 2MASS} images, there is no indication of
  any object at the optical/NIR coordinates of \vsx}.  We have,
however, found a detection of \vsx\ in the UKIDSS\footnotemark
\footnotetext{UKIDSS is the UKIRT Infrared Deep Sky Survey.  See
  http://www.UKIDSS.org.}  database (Lawrence et al.  2007) from UT
2006 June 10; the only wavelength at which it is brighter than the
{\it 2MASS} detection limits is K' where it has a magnitude of
$\sim$13.3.  A comparison of the {\it 2MASS} and UKIDSS K' images
(taken in 2000 and 2006, respectively) is shown in
Figure~\ref{2mass-ukidss}.  Magnier et al.  (1999) included \vsx\ in a
catalog of young stellar objects (YSOs) and noted that at far-IR
wavelengths it had a flat spectrum suggestive of an object
transitioning from the protostellar to T~Tauri evolutionary phase.
Finally, Covey et al.  (2011) present NIR photometric observations
from five epochs spanning the period July 2010 to September 2010.

\begin{figure}[t] 
\begin{center}
\includegraphics*[angle=270,scale=0.7]{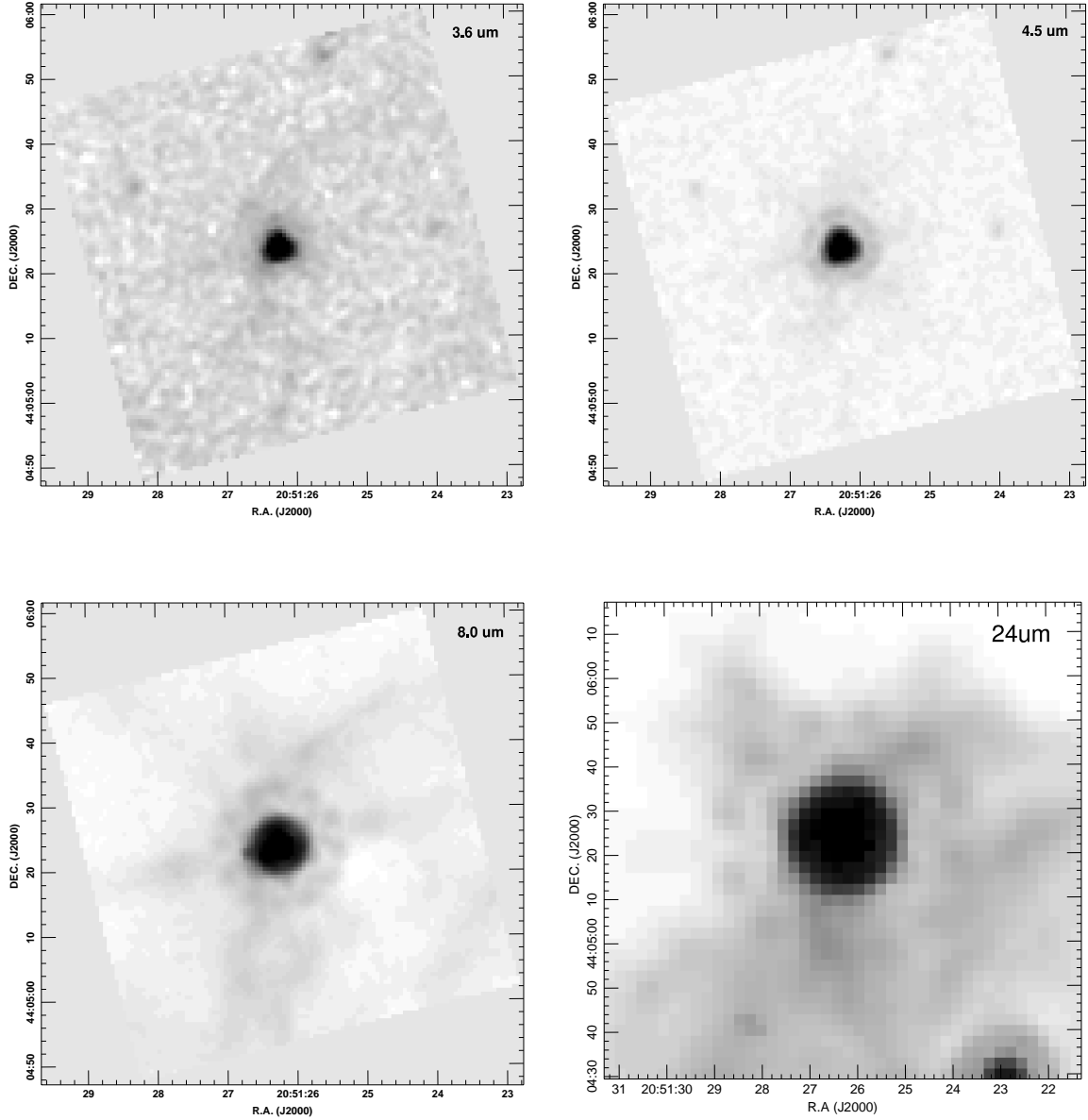} 
\caption{{\it Spitzer} IRAC and MIPS images of \vsx.  These 
images are $\sim$72$''$ square and are from the IRAC band 1 [3.6~$\mu$m] 
(top-left),  the IRAC band 2 [4.5~$\mu$m] (top-right), the IRAC band 4 
[8.0~$\mu$m] (bottom-left), the MIPS band 1 [24~$\mu$m] (bottom-right) 
images.  \vsx\ is the bright point-source at the center of each image.  
The IRAC images were acquired on UT 2006 August 9 while the MIPS image 
was taken on UT 2006 June 10.
\label{vsxspitims}}

\end{center}
\end{figure}

The {\it AKARI/IRC} Point Source Catalog (Murakami et al. 2007)
contains a source within 6$''$ of the optical/NIR coordinates with 9
and 18~$\mu$m fluxes of 2.0 and 2.9~Jy, respectively.  The star was
not detected in {\it AKARI/FIS} observations. {\it Spitzer} did,
however, detect \vsx\ in all four IRAC bands and with MIPS at 24 and
70~$\mu$m.  {\it Spitzer} images, centered on \vsx, are shown in
Figure~\ref{vsxspitims}.  {\it MSX6C} also detected \vsx\ at infrared
wavelengths from 8.28 to 21.34~$\mu$m.

The {\it IRAS} source associated with \vsx\ has Point-Source Catalog
(PSC) fluxes at 12, 25, 60, and 100~$\mu$m fluxes of 3.4, 6.6, 28.2,
and 58.3~Jy, respectively.  The precessed {\it IRAS} coordinates are
within 0$\farcs$5 of the optical/NIR position and therefore well
within the 16$''$ {\it IRAS} error ellipse.  No obvious
optical/infrared sources are in the immediate vicinity of \vsx, making
the association of optical and {\it IRAS} sources firm.  One
interesting point, however, is that the {\it IRAS} 60~$\mu$m flux
(taken in 1983) is 5$\times$ larger than the 60~$\mu$m {\it
  Spitzer/MIPS} flux (taken in 2006/2007).

\subsection{At mm Wavelengths}
The {\it Bolocam} Galactic Plane Survey (GPS) contains 1.1~mm images
of the region containing \vsx\ and lists a source within 4$''$ of its
coordinates (GPS designation G084.466-00.135).  The flux quoted in the
GPS catalog is 148$\pm$74~mJy and since the next closest mm source is
46$''$ away, we adopt this value for its 1.1~mm flux.

\section{RESULTS}
\subsection{Optical Imaging}
The optical images of \vsx\ presented in Figure~\ref{ims} show a
pre-outburst 1997 H$\alpha$ image taken from Bally \& Reipurth (2003,
left) and our post-outburst Gemini GMOS r' image (right) taken on UT
2010 September 5.  \vsx, located at the image centers, is very much
brighter post-outburst.  No obvious optical nebulosity is seen around
\vsx, even in its photometrically elevated state.  Nor was one found
in the R-band images of Covey et al. (2011).  For comparison, we note
that all classical FUors and FUor-like objects have been found to
exhibit nebulous structure in outburst which often takes the form of
curving tails extending from the young star (Goodrich 1985).  EXors do
not typically possess associated nebulosity.

Due to source saturation, we could not obtain photometry from our
optical images, however, we note that at the time of the GMOS
observations, the American Association of Variable Star Observers
(AAVSO) database shows the source having an optical V-band magnitude
of 15.1$\pm$0.1.  In Figure~\ref{lcs} we show the AAVSO V-band light
curve of \vsx.  In addition to the AAVSO data\footnotemark
\footnotetext{We refer the reader to the AAVSO web site at
  http://www.aavso.org for access to the V-band photometry shown in
  Figure~\ref{lcs}.}, we have also included the 'unfiltered'
photometry from Itagaki (2010, blue downward-facing arrows) and the
R-band data from Covey et al. (2011, grey filled circles).  The
Itagaki (2010) photometry serves as an indication of the optical
progression of the rise from quiescence to outburst and since these
data are in unfiltered optical light, they are upper limits.  The
Covey et al. (2011) data indicates both the level of variability
present over the time period shown.

\subsection{Near-Infrared Photometry}
We can place \vsx\ in NIR color-color (henceforth c-c) diagrams and
thereby investigate the level and associated variability of overlying
extinction and intrinsic thermal excess emission.  One note of caution
is, however, warranted.  If \vsx\ is a Class~I protostar, its NIR
fluxes may be significantly modified by circumstellar scattering.
This was studied in detail by Whitney, Kenyon, \& Gomez (1997) who,
using simple extinction+scattering models, discussed the effect of
scattering on emergent flux.  Below we assume that the direct flux
from \vsx\ is not dominated by scattering but that it is true,
reddened stellar/disk flux.  Our results are likely not applicable if
scattered flux forms the major component of the observed NIR signal.

Figure~\ref{ccs}a shows on the a NIR J--H vs. H--K (henceforth JHK)
while Figure~\ref{ccs}b shows the H--K vs.  K--L (henceforth HKL) c-c
diagram.  Some of the 2~and~3~$\mu$m photometric data included were
taken using filters with slightly different central wavelengths and
bandpasses (e.g. K and K', and L and L'), however, within the
associated uncertainties (typically $\sim$10\% on photometry,
$\sim$14\% on colors) we consider the slight differences in
magnitudes/fluxes between filters unimportant for the analysis below.
Our new NIR photometric values are detailed in Table~\ref{nirphot}.

In the JHK c-c diagram (panel a), points \#1, \#4, and \#8 shows the
location of \vsx\ at three different epochs i.e. the UKIDSS (JHK')
observations (UT 2006 June 10) and our two Gemini/NIRI (JHK'L')
observations (UT 2010 September 5 and November 25).  In our
Gemini/NIRI data from September 5 (the closest observation to the peak
optical brightness), \vsx\ has a K' magnitude of 8.12$\pm$0.1 while
approximately 3 months later (November 25) it had faded to
9.63$\pm$0.1.  With a pre-outburst UKIDSS K' magnitude of 13.15, the
source had therefore brightened at 2~$\mu$m by $\sim$5~mags between
2006 (\#1) and 2010 (\#4).  This is likely the result of a combination
of a decrease in 2~$\mu$m thermal excess and/or a reduction in
overlying visual extinction.\footnotemark \footnotetext{An alternative
  may be that the disk is hotter at maximum and optically thick gas
  interior of the dust sublimation radius adds NIR continuum flux.} In
addition to points \#1, \#4, and \#8, we have also included the J, H,
and K photometry from Covey et al.  (2011), points \#2, \#3, and \#5,
\#6, and \#7, in chronological order.  The trend from the earliest
photometry (\#1) to the observation just after maximum brightness
(\#4) is approximately linear with both J--H and H--K decreasing with
time.  This progression is close to being along a reddening vector
suggesting that the change is due to a reduction in overlying
extinction.  In Figure~\ref{ccs}a we show reddening vectors (dashed
lines) using the extinction law defined by Robitaille et al.  (2007)
and used in their young star spectral energy distribution (SED)
models.  These extend from the extremes of the zero-age main
sequence/giant branch and from the end of the CTTS locus defined by
Greene \& Meyer (1995), the latter extends between points labeled A
and B.  It may be the case that the linear change in location of \vsx\
in the JHK c-c diagram is the result of the presence of a somewhat
different reddening law than that shown in Figure~\ref{ccs}a.  Perhaps
more plausibly, the variation in JHK colors could be caused by a
combination of changes in overlying extinction and K-band thermal
excess.

To explore the first of these two options, we plot a selection of
reddening laws together with the \vsx\ data in Figure~\ref{ccs2} with
the sequence of JHK observation of \vsx\ shown as filled black
circles.  Point \#1, however, is outside the area plotted but is
located on the dashed green line (see Figure~\ref{ccs}a for details).
All reddening vectors, together with the dotted line extending to
point \#1, originate at the red end of the CTTS locus (point B, again,
see Figure~\ref{ccs}a).  We have plotted four different reddening laws
defined in the literature and often used in young star JHK diagrams.
All vectors represent A$_V$=10~mags.  The solid grey lines show the
Cardelli, Clayton, and Mathis (1989, henceforth CCM) reddening vectors
for four different values of R, the ratio of total to selective
extinction.  The values shown are for R~=~2, 3.1, 5.0, and 7.0.  R=3.1
is the typical interstellar reddening value while R=5.0 is often
quoted for extinction towards young stars in dense dark clouds.  As
can be seen, all four CCM reddening vectors have very similar slopes
with the major effect of changing the value of R being the length of
the A$_V$=10~mags reddening vector.  Close to the CCM vectors lies a
generic reddening law in which the wavelength dependence of reddening
is proportional to the scaled wavelength (with respect to the value at
0.55~$\mu$m) to the power of $\alpha$=--1.6 (red dashed line).
Reddening from this relationship closely matches the CCM reddening
laws.  The single-dotted-dash line (blue) shows the Rieke \& Lebofsky
(1985) reddening law defined using observations of the galactic center
while the triple-dotted-dash line (purple) shows the law adopted by
Robitaille et al. (2007) in the generation of their SED models.  The
latter two reddening laws produce somewhat different slopes with
respect to the CMM and $\alpha$=--1.6 laws.  However, none of these
reddening laws extend through the sequence of \vsx\ colors.  What we
can conclude from Figure~\ref{ccs2} is that neither changes in the
adopted reddening law nor in the value of R used would result in the
colors variations observed in \vsx.

The location of the three Covey et al. (2011) JHK observations,
obtained between our two Gemini/NIRI datasets (i.e. points \#5--7),
indicate that within a few days of our acquisition of the data that
gave rise to point \#4 the source's JHK colors began to reverse the
progression observed (from points \#1 to \#4).  A similar behavior was
seen during the 2003--2006 eruption of V1647~Ori (Aspin, Beck, \&
Reipurth 2008) where it was attributed to a variation of overlying
extinction of around 10~mags.  However, in the case of the color
variations of V1647~Ori, they occurred directly along a Rieke \&
Lebofsky (1985) reddening vector.  Our most recent observation, point
\#8, shows that this reversal continued and at the end of November
2010 the source's JHK colors were similar to what they were in July
2010 (points \#2 and \#3) although in November 2010, \vsx\ had a K'
magnitude of 9.63$\pm$0.1 while in July 2010 Covey et al. found a
value of 8.45$\pm$0.02.  As an exercise, if we attribute the total
change in JHK colors from point \#2 to point \#4 to changes in A$_V$,
then the amount of overlying visual extinction would have to have
reduced by $\sim$7~mags.  If we also attribute the extremely large
colors in June 2006 (point \#1) to reddening, then the change of A$_V$
from \#1 to \#4 would amount to $\sim$20~mags.

In our HKL c-c diagram (Figure~\ref{ccs}b, right) there exists
only two epochs of data.  Point \#4 (2010 September 5) shows \vsx\ to
possess a significant 3~$\mu$m thermal excess suggesting that even
though the outburst had already occurred, a considerable amount of
heated dust must still be present in the circumstellar environment
close to the star.  Approximately three months later (UT 2010
November~25, point \#8), the 3~$\mu$m thermal excess had grown even
larger implying a continuing build-up of dust close enough to the
star/accretion region to be heated to temperatures of $\sim$1000~K.
This dust, however, must be far enough so as not to be sublimated at
temperatures of $\sim$1500~K.  Tuthill et al. (2001), and subsequently
Lykou et al.  (2010), suggested that the distance from a star where
dust sublimates can be estimated using the equation

\begin{equation}
  r_{sub} = 1.1\times(T_{sub}/1500~K)^{-2}\times(L_{bol}/1000~L_\odot)^{0.5}
\end{equation}

where r$_{sub}$ is the sublimation radius in AU, T$_{sub}$ is the
sublimation temperature in K, and L$_{bol}$ is the system bolometric
luminosity in L$_{\odot}$.  Assuming T$_{sub}$~=~1500~K and a
pre-outburst L$_{bol}$ of $\sim$35~L$_{\odot}$ (Bally \& Reipurth
1997; Covey et al. 2011), we obtain a value for r$_{sub}$ of
$\sim$0.2~AU.  In its elevated state, Covey et al. (2010) suggest that
the bolometric luminosity increases by a factor 100. This would result
in r$_{sub}$ increasing by a factor 10 to $\sim$2~AU.  Therefore, we
suggest that the dust content of the disk region between 0.2 and 2~AU
must have been accreted or sublimated during the outburst.

\subsection{Spitzer Photometry}

Similarly, c-c plots can be made from {\it Spitzer} IRAC and MIPS
photometry.  Such diagrams are useful for assigning evolutionary
classes to young stars since in them Class~I protostars, Class~II
classical T~Tauri stars (CTTSs), Class~III weak-line T~Tauri stars,
and main sequence stars are well separated (Allen et al. 2002).  We
have extracted photometric magnitudes from the {\it Spitzer} IRAC and
MIPS images using the IDL routine {\it photvis} (v1.1, Gutermuth et al.
2004, 2008\footnotemark \footnotetext{We have used the same aperture
  sizes and photometric zeropoints as in the aforementioned papers.}).
These values are given in Table~\ref{sedphot}.  In the IRAC
[3.6]--[4.5] vs.  [5.6]--[8.0] plot (Figure~\ref{spitccs}a) \vsx\
(point \#1) lies well within the Class~I regime.  In the IRAC/MIPS
[3.6]--[4.5] vs. [8.0]--[24] plot (Figure~\ref{spitccs}b), the
source lies on the Class~I--II boundary.  In these Figures we also
show the locations of several FUors for comparative purposes (points
\#2 through \#6) together with the location of \vsx\ using the {\it
  Spitzer} fluxes given in Covey et al.  (2011, point \#7).  What we
can conclude from these plots is that the location of \vsx\ is
consistent with it being a Class~I protostar.  We note that in the
[3.6]--[4.5] vs.  [5.6]--[8.0] plot, \vsx\ lies closest to the
FUor-like object L1551~IRS5 (point
\#5)\footnotemark\footnotetext{L1551~IRS5 was saturated in the {\it
    Spitzer} MIPS 24~$\mu$m images and so our photometry is that given
  at 24~$\mu$m in Osorio et al.  (2003).}. We are uncertain as to why
the Covey et al. (2011) [3.6]--[4.5] vs.  [8.0]--[24] \vsx\ point is
displaced from our measurement since both use the same datasets.
Perhaps this indicates that the uncertainties on the extracted
24~$\mu$m photometry are somewhat larger than the error bars shown.

\subsection{Optical Spectroscopy}
\subsubsection{The UT 2010 September 5 GMOS-N Spectrum}
Our GMOS-N optical spectrum of \vsx\ from UT 2010 September 5 is shown
in Figure~\ref{full} (top).  It has a signal-to-noise of $\sim$200 and
was acquired close to maximum optical brightness.  The spectrum is
heavily veiled and, hence, no photospheric absorption features are
visible.  \vsx\ shows a wealth of emission lines including H$\alpha$
(6563~\AA), \ion{Ca}{2} (8498, 8542, and 8662~\AA), and numerous
neutral and singly ionized Fe lines.  Both \ion{K}{1} (7665 and
7699~\AA), and \ion{O}{1} (7773 and 8446~\AA) are, however, in
absorption.  We note that \ion{Li}{1} is not detected.  The complete
list of emission and absorption lines with equivalent widths, EW,
$\ge$1~\AA, together with flux values, are given in
Table~\ref{fluxes}.  Figure~\ref{full} (bottom) also shows a
comparison of this optical spectrum with that of an M6~V dwarf.  As
Covey et al. (2011) discovered, and we confirm here, the TiO bands,
found in absorption in late-type stars, are in emission in \vsx.  The
spectrum of the M6~V dwarf is used here as an example of a late-type
star showing strong TiO band absorption and we have marked the extent
of the TiO emission in \vsx\ by horizontal lines above the M6~V
spectrum.

Although the resolving power of our GMOS spectra is only {\it
  R}$\sim$1200, we can nevertheless obtain some insight into the
velocity structure present in various spectral features via line
profile fitting.  We note that an unresolved arc lamp line using the
B600 grating and 0$\farcs$75 slit has a FWHM$\sim$130~km~s$^{-1}$ and,
hence, profile fitting should conservatively give reliable velocity
information at the $\sim$20\% level or $\sim\pm$30~km~s$^{-1}$.  It is
this value we adopt for the uncertainty on quoted velocities.

The H$\alpha$ emission feature shows associated blue-shifted
absorption out to velocities of $\sim$--450~km~s$^{-1}$, similar to
that found by Covey et al. (2011).  The \ion{Ca}{2} triplet emission
lines possess weak blue-shifted absorption components out to
velocities of $\sim$--300~km~s$^{-1}$.  The absorption components on
H$\alpha$ and \ion{Ca}{2} can be better seen in the expanded views
shown in Figure~\ref{vhacaii}.  H$\alpha$ has an EW of --16.3~\AA\ but
in reality the value will be somewhat larger (i.e.  more negative) due
to the presence of the associated absorption components.  The
(deconvolved) full-width 10\% intensity (FW10\%) of H$\alpha$ is
550~km~s$^{-1}$ while the location of the minimum of the absorption
component is --210~km~s$^{-1}$.  The peak of the H$\alpha$ emission is
close to zero velocity as it is for the \ion{Fe}{1} lines and the
\ion{Ca}{2} triplet lines.  The velocity of the minimum blue-shifted
absorption associated with the \ion{Ca}{2} triplet lines is
$\sim$--190~km~s$^{-1}$ and, within the associated uncertainties, is
the same as on H$\alpha$.

\subsubsection{The UT 2010 November 25 GMOS-N Spectrum}
Optical spectroscopy of \vsx\ was also obtained on Gemini-North
using GMOS on UT 2010 November 25.  The same instrumental setup was
used as for the September 5 spectrum.  Figure~\ref{full2} shows this
spectrum in the same format as the earlier spectrum.  At the time of
acquisition, \vsx\ had optically faded by 4.5 ~mags and had an
R-band magnitude of 18$\pm$0.2 (Covey 2010, private communication).

There are several differences between the optical spectra taken on UT
2010 September 5 and November 25, specifically, the TiO emission bands
have slightly changed shape (as was also seen by Covey et al. 2011),
H$\alpha$ has lost its deep blue-shifted absorption component, and
several other spectral lines have changed strength or have
appeared/disappeared.  The fading of \vsx\ from R$\sim$13.5 to
18.0 between UT 2010 September~5 and November~25 would correspond to
an EW increase for constant line emission of $\sim$60$\times$.  With
this in mind, we find that:

\begin{itemize}

\item As noted above, H$\alpha$ has lost its deep blue-shifted
  absorption component.  Absorption appears to be present nonetheless
  but at lower velocities.  One possible scenario is that H$\alpha$
  emission has filled in the blue-shifted absorption.  The EW of the
  H$\alpha$ emission is --27.6~\AA as opposed to --16.3~\AA\ on
  September 5.  For constant H$\alpha$ emission, the change in
  continuum level between the two dates would result in an H$\alpha$
  EW of $\sim$1000~\AA.  We can conclude that the H$\alpha$ emission
  has declined in intensity by a factor $\sim$35 between the two
  observations.  In addition, the FW10\% intensity of H$\alpha$ has
  declined from 550~km~s$^{-1}$ to 300~km~s$^{-1}$.

\item The [\ion{O}{1}] lines at 6300 and 6363~\AA\ have become much
  more prominent and rival H$\alpha$.  Their EW changed from --1.3 and
  --0.6~\AA\ on September 5 to --38.9 and --13.3~\AA\ on UT 2010
  November 25.  This increase in EW is a factor of about 30 suggesting
  that the [\ion{O}{1}] emission declined by around a factor 2 between
  the two dates.  However, the \ion{O}{1} absorption lines at 7774 and
  8446~\AA\ have weakened by $\sim$factor 2 over the same time period.
  
\item The reappearance of [\ion{S}{2}] lines at 6716 and 6731~\AA,
  which were absent in the September 5 spectrum yet present in the
  optical spectra of Covey et al. (2011), suggests that shocks have
  reappeared in the intervening period.  The ratio of the EWs of the
  two [\ion{S}{2}] emission lines (6716/6731) is $\sim$0.6.  Since the
  production of these lines and their flux ratio are only weakly
  dependent on temperature and more strongly dependent on density, we
  have used the IRAF program {\tt nebular.temden} to calculate the
  electron density, {\it n$_e$}, necessary to produce this ratio.  We
  find that a high density environment is needed with an {\it
    n$_e$}$~\sim$4000~cm$^{-3}$ for an electron temperature of
  $\sim$10$^4$~K.  We note that the critical density for [\ion{S}{2}]
  is $\sim$5$\times$ larger than this value and that typical electron
  densities in Herbig-Haro shocks is $\sim$1000~cm$^{-3}$.

\item The disappearance of blue-shifted absorption on the \ion{Ca}{2}
  triplet line at 8498~\AA\ is peculiar especially since it is still
  present and well detected on the 8543 and 8662~\AA\ \ion{Ca}{2}
  triplet lines.  On September~5, this line, however, had the weakest
  blue-shifted absorption of the three lines (see
  Figure~\ref{vhacaii}).  The velocity of the minimum in the
  blue-shifted absorption has declined from --190~km~s$^{-1}$ to
  --125~km~s$^{-1}$, a proportionally similar velocity change as the
  blue-shifted absorption on H$\alpha$.  Interestingly, the EW of the
  \ion{Ca}{2} triplet lines are similar on the two observing dates.
  This implies that as the broad-band R flux declined the \ion{Ca}{2}
  emission declined by a similar amount.

\item The \ion{K}{1} absorption lines at 7665 and 7699~\AA\ behave in
  a similar manner to the \ion{Ca}{2} triplet lines in that their EW
  remained approximately constant in the September and November
  spectra.

\end{itemize}

\subsection{NIR Spectroscopy}
\subsubsection{The UT 2010 October 2/3 NIFS Spectra}
The NIFS J, H, and K spectra of \vsx\ taken on UT 2010 October 2 (K
band) and 3 (J and H bands) are shown in Figure~\ref{compv-oct}.  The
full spectral range observed is shown in the top-left panel with the
J, H, and K bands shown in the top-right, bottom-left, and
bottom-right panels, respectively.  For comparison we also show the
prototypical EXor eruptive variable EX~Lup.\footnotemark
\footnotetext{The EX~Lup spectrum is from the SpeX spectrograph at the
  NASA IRTF telescope and has a lower spectral resolution of
  $\sim$1200.} Apart from the continuum slope, which increases with
wavelength in \vsx\ and decreases with wavelength in EX~Lup, almost
all features observed in EX~Lup are reproduced in \vsx\ and are
totally dissimilar to those seen in FUors such as FU~Ori.
Specifically, the CO overtone bandheads are strongly in emission as
are Br$\gamma$ and Pa$\beta$.  Other \ion{H}{1} emission lines are
also present as are emission lines from \ion{Mg}{1}, \ion{Na}{1}, and
\ion{Fe}{1}\footnotemark \footnotetext{The forest of emission lines
  from 1.15 to 1.22~$\mu$m are mostly from neutral Fe.}.  We
additionally note that the broad water vapor bands at 1.4~$\mu$m and
1.8~$\mu$m are weakly in absorption in EX~Lup while in \vsx\ they
appear weakly in emission (as indicated by the small increase in
continuum slope near those wavelengths).

The observed Pa$\beta$ and Br$\gamma$ line fluxes in the UT 2010
October 2/3 NIR spectra of \vsx\ are 1.1$\times$10$^{-13}$
(EW=--11.8~\AA) and
9.0$\times$10$^{-14}$~ergs~cm$^{-2}$~s$^{-1}$~\AA$^{-1}$
(EW=--5.1~\AA), respectively.  These correspond to line luminosities
of $\sim$0.001~L$_{\odot}$ in both Pa$\beta$ and Br$\gamma$ assuming
A$_V$=0, d=600~pc, and that the line flux is emitted isotropically.  A
complete list of spectral features present in the NIR spectra of \vsx\
is given in Table~\ref{fluxes2}.

Since \vsx\ exhibits CO overtone bandhead emission, we can investigate
the characteristics of the CO emission region using simple numerical
models.  We follow our previous CO modeling in Aspin et al.  (2010)
where we studied the bandhead structure in EX~Lup during its 2008
extreme eruption.  The model we used is based on that of Dent \&
Geballe (1991) and Carr \& Tokunaga (1992) and calculates the CO
bandhead emission from an isothermal slab of emitting molecular gas
characterized by a CO temperature, T$_{CO}$, and optical depth in the
CO bandhead, $\tau_{CO}$, and a gas velocity profile.  The velocity
profiles we have used are either Gaussian of FWHM {\it v$_g$} or that
of a Keplerian rotating gas disk with velocities defined by the mass
of the star, M$_*$, the inner and outer radii of the CO gas disk, {\it
  r$_{in}$} and {\it r$_{out}$}, respectively.  Although this model is
relatively simple, its ability to produce good matches to CO bandhead
emission spectra and its small number of free-parameters lend
credence to the uniqueness of the values obtained and the values
themselves.

Prior to presenting the results of the CO bandhead modeling, it is
informative to comment on the effect of changing the various free
parameters used on the model spectrum.  This is shown in
Figure~\ref{co2}.  In panel {\it (a)} we plot a model (blue) using the
free-parameters {\it v$_g$}~=~55~km~s$^{-1}$, T$_{CO}$~=~3200~K, and
$\tau_{CO}$~=~1.8.  In the subsequent panels we vary one of the
free-parameters of the model to demonstrate the changes that occur.
In panels {\it (b)} and {\it (c, red)} we change {\it v$_g$} to 100
and 25~km~s$^{-1}$, respectively.  Note that the bandhead slope
changes as does the visibility of the higher order transitions
(resulting in the sinusoidal structure).  In panels {\it (d)} and {\it
  (e, green)} we change T$_{CO}$ from 1500 to 4000~K, respectively.
The effect of changing T$_{CO}$ is to raise or lower the level of the
emission longward of the bandhead wavelength, specifically, {\it i)}
at low T$_{CO}$ values (e.g.  1500~K) the higher order CO transitions
are stronger and hence the bandhead emission remains strong to longer
wavelength, and {\it ii)} at high T$_{CO}$ values the opposite is true
and the longer wavelength CO bandhead emission is suppressed.  In
panels {\it (f)} and {\it (g, purple)} we change $\tau_{CO}$ from an
optically thin $\tau_{CO}$~=~0.01 to optically thick
($\tau_{CO}$~=~3.6).  This changes the emission strength at
wavelengths longward of the bandhead and, in some ways, mimics the
effect of changing T$_{CO}$.  If we next consider a disk velocity
profile, as shown in panels {\it (h, pink)}, the so-called {\it 'blue
  hump'}\footnotemark\footnotetext{The 'blue hump' is the name given
  to the enhanced emission on the blue side of the CO bandhead limit
  at 2.2935~$\mu$m).} appears and the shape of the bandhead changes
significantly from that occurring when using a simple Gaussian
velocity profile.  The disk velocity profile used in panel {\it (h)}
we had M$_*$~=~0.6~M$_{\odot}$, T$_{CO}$~=~2500~K, $\tau_{CO}~<$~0.1,
and {\it r$_{in}$} and {\it r$_{out}$}~=~1.2 and
2.0$\times$10$^{10}$~m, respectively.  These were the values used for
the best-fit model to the EX~Lup spectrum shown in
Figure~\ref{co1-oct}.  If we obscure part of the disk velocity
profile, as might be the case in a real star--disk system, we obtain
the CO bandhead profiles shown in panels {\it (i)} and {\it (j,
  pink)}.  In panel {\it (i)} we have removed all blue-shifted
velocities from the disk velocity profile and the resultant spectrum
shows a significantly red-shifted bandhead peak and peculiar structure
in the higher CO transitions.  If we remove all red-shifted velocities
from the disk velocity profile (panel j) we see a CO bandhead with a
'blue hump' and truncated emission longward of 2.2935~$\mu$m. Clearly
intermediate obscuration between these two extremes will produce an
intermediate CO profile; such spectral peculiarities could be
considered as diagnostics of a partly obscured gas disk around a young
star.

For \vsx, we have investigated a wide range of values for the above
free parameters for both a Gaussian and disk velocity profile.
Specifically, these are {\it v$_g$}~=~0--150~km~s$^{-1}$,
T$_{CO}$~=~1500--4500~K, $\tau_{CO}$~=~0.01--10,
M$_*$~=~0.1--10~M$_{\odot}$, and {\it r$_{in}$} and {\it
  r$_{out}$}~=~10$^9$--10$^{11}$~m with {\it r$_{out}$}~$>$~{\it
  r$_{in}$}.  The best-fit model to the observed CO bandhead profile
required a Gaussian velocity profile with {\it
  v$_g$}~=~55$\pm$10~km~s$^{-1}$, T$_{CO}$~=~3200$\pm$200~K, and
$\tau_{CO}$~=~1.8$\pm$0.1.  This model (in blue) is shown overlaid on
the observed profile (in red) in Figure~\ref{co1-oct}.  For
comparison, we also show the observed (black) and best-fit model
profile (grey) for EX~Lup from Aspin et al.  (2010).  Here, a disk
profile was required to fit the observations.  Although the spectral
resolution of the EX~Lup spectrum was considerably higher than that of
our Gemini/NIFS spectrum (Keck~II/NIRSPEC with R$\sim$18,000 as
opposed to Gemini/NIFS with R$\sim$5,000), a version of the EX~Lup
spectrum binned to match our \vsx\ resolution (green) shows that the
observed 'blue-hump' structure, indicative of a disk velocity profile,
is still clearly present.  We conclude, therefore, that a disk
velocity profile is not required to reproduce the observed CO bandhead
profile in \vsx, rather, a simple Gaussian profile is sufficient.
This suggests that the CO emission arises from gas with a simple
Gaussian velocity profile, perhaps in accretion funnels or in regions
with little radial velocity component such as the inner edge of an
accretion disk viewed pole-on.

\subsubsection{The UT 2010 November 26/27 NIFS Spectra}
Second epoch NIR NIFS spectra of \vsx\ was obtained on UT November 26
(K band) and 27 (H band)\footnotemark \footnotetext{No J band spectrum
  was acquired due to poor weather conditions and instrument
  problems.}.  These spectra are presented in Figure~\ref{compv-nov}.
As in Figure~\ref{compv-oct}, top-left is the full 1--2.5~$\mu$m
spectrum of \vsx\ together with the same EX~Lup spectrum.  Bottom-left
is a closer view of the H-band spectra and bottom-right is the same
for the K-band.  Instead of the J-band spectrum at the top-right, we
now show a dereddened version of the full 1--2.5~$\mu$m spectrum.  The
dereddening used an extinction of A$_V$=20~mags and shows that such a
value for optical extinction produces a spectral slope that more
resembles that of EX~Lup (which has an intrinsic A$_V$ of around
0--1~mags).  The EW, line flux, and continuum values for this spectrum
are also shown in Table~\ref{fluxes2}.

The main differences between the UT 2010 October 2/3 and November
26/27 NIR spectra of \vsx\ are:

\begin{itemize}
\item The continuum slope is now considerable redder than previously
  seen.  This is consistent with the significant change in NIR colors
  between the two epochs and related above. 

\item The water vapor bands exemplified by the 1.7--2.2~$\mu$m
  feature, are more strongly in emission than previously observed. 

\item The CO overtone bandheads remain in emission although their
  strength has declined by a factor $\sim$4 during the intervening
  period.  The profile of the v=2-0 CO bandhead at 2.294~$\mu$m is
  similar to that seen in October, however, the spectral structure
  observed is less well defined and the slope of the bandhead itself
  is slightly shallower.  Fitting our CO bandhead emission model to
  the UT 2010 November 26 data produces a best fit as shown in
  Figure~\ref{co1-nov}.  This model requires {\it
    v$_g$}=85$\pm$10~km~s$^{-1}$, T$_{CO}$=4200$\pm$200~K, and
  $\tau_{CO}$=2.0$\pm$0.1 to reproduce the observations.  This
  represents increases in Gaussian velocity, CO temperature, and
  optical depth with respect to the values found for the best fit to
  the UT 2010 October 2/3 data.  Since the same instrument and
  instrumental setup was used for both sets of observations, we
  consider these changes to be significant at the level determined by
  the associated uncertainties.

\item Molecular hydrogen (H$_2$) emission, in the form of the v=1-0
  S(1) line at 2.122~$\mu$m, has become more apparent although its
  intrinsic strength has remained approximately constant at
  $\sim$1.4$\times$10$^{-14}$~ergs~cm$^{-2}$~s$^{-1}$.  [\ion{Fe}{2}]
  emission at 1.644~$\mu$m is now detected albeit at a weak level
  ($\sim$3.6$\times$10$^{-15}$~ergs~cm$^{-2}$~s$^{-1}$).  Both these
  lines are typically associated with the generation of shocks.

\item The Br$\gamma$ line flux decreased by a factor $\sim$4 between
  UT 2010 October 2/3 and November 26/27.  This is very similar to the
  reduction in both the CO bandhead emission and the \ion{Na}{1}
  emission at 2.208~$\mu$m.  The profile of Br$\gamma$ is very similar
  in the two spectra with a (deconvolved) FWHM$\sim$200~km~s$^{-1}$.
  The line luminosity in Br$\gamma$ has declined from
  0.001~L$_{\odot}$ on UT 2010 October 2 to 0.0003~L$_{\odot}$ on
  UT 2010 November 26, again assuming A$_V$=0 and d=600~pc.

\end{itemize}

\subsubsection{A Comparison to the pre-October 2010 NIR Spectra}
Covey et al. (2011) presented NIR spectra of \vsx\ taken on UT 2010
July 14 and 18.  On these dates the star had an optical brightness of
R$_{PTF}$\footnotemark \footnotetext{PTF stands for the Palomar
  Transient Factory.  Definition of the R$_{PTF}$ filter is given in
  Covey et al. (2011).} $\sim$15.5 and 16, respectively.  These values
are around 2 and 2.5~mags fainter (on the R$_{PTF}$ magnitude scale)
than at maximum light which occurred in early September 2010. In
quiescence (e.g. in July 2009), \vsx\ had an R$_{PTF}$ brightness of
$\sim$19.

We find, in comparison to our subsequent spectra, that:

\begin{itemize}

\item On UT 2010 July 14 and 18, \vsx\ possessed Br$\gamma$ line
  fluxes of 1.4 and 2.9$\times$10$^{-14}$~ergs~cm$^{-2}$~s$^{-1}$,
  respectively.  This compares to 9 and
  2.3$\times$10$^{-14}$~ergs~cm$^{-2}$~s$^{-1}$ on UT October 2 and
  November 26, respectively.  If we equate Br$\gamma$ line flux to
  accretion then we can conclude that in early October 2010,
  \vsx\ had an accretion rate some $\sim$3--5$\times$ of that in
  July and November 2010.

\item The v=1-0~S(1) H$_2$ emission line at 2.122~$\mu$m appears to be
  relatively constant over the period UT 2010 July 14 to November 26.
  The line flux measured by Covey et al. (2011) and in our spectra was
  1, 1.7, 1.3, and 1.5$\times$10$^{-14}$~ergs~cm$^{-2}$~s$^{-1}$, on
  UT 2010 July 14 and 18, September 5, and November 26, respectively.
  The v=2-1~S(1) H$_2$ line at 2.2477~$\mu$m line is detected at three
  of the four epochs, specifically, July 14 and 18, and November 26.
  Since it is significantly weaker than the v=1-0~S(1) line this is
  understandable due to the v=1-0~S(1) line flux being the smallest on
  UT 2010 October 2.  Beck et al. (2007) considered the ratio of
  certain H$_2$ lines for different sources of excitation\footnotemark
  \footnotetext{Fluorescent excitation ratio calculated by Black \&
    van~Dishoeck (1987).  Shock excitation ratios calculated by Smith
    (1995).}  as a diagnostic for the excitation mechanism in effect.
  The ratio of the v=2-1~S(1)/v=1-0~S(1) was found to be 0.55, 0.05,
  and 0.24 for UV pumped fluorescent excitation, continuous C-type
  shock excitation\footnotemark \footnotetext{C-shocks occur in
    regions of strong magnetic field which buffer the higher velocity
    ($\sim$50~km~s$^{-1}$) continuous shock from dissociating the
    H$_2$ molecules (Draine \& McKee 1993).}, and jump J-type shock
  excitation\footnotemark \footnotetext{J-shocks show a discontinuous
    change in density, temperature, and velocity across the shock
    front and are typically of lower velocity ($<$35~km~s$^{-1}$)
    which does not dissociate the H$_2$ molecule (Draine \& McKee
    1993).}, respectively.  On the three dates the v=2-1~S(1) line was
  detected, the above ratio was 0.24, 0.33, and 0.19 suggesting that
  the H$_2$ excitation mechanism is likely to be J-type shocks.

\item The flux ratio of Br$\gamma$ emission to \ion{Na}{1}
  2.206~$\mu$m emission was found by Lorenzetti et al. (2009) and
  Antoniucci et al. (2008) to be dependent on the type of young star
  being observed.  Covey et al. (2011), used the ratio of these lines
  to show that for \vsx, the value was $\sim$2 suggesting that the
  emission in the source was more typical of that from an EXor/jet
  driving source.  Our two epochs of NIR spectroscopy give a ratio of
  2.4 and are hence consistent with the values of Covey et al.

\end{itemize}

\section{Discussion}
\subsection{Pre-Outburst Spectral Energy Distribution}

Using the photometry in Table~\ref{sedphot} we can construct a
pre-outburst SED of \vsx.  This is shown graphically in
Figure~\ref{seds}.  We have used the Robitaille et al. (2006, 2007)
models and on-line SED fitting tool\footnotemark\footnotetext{At URL:
  http://caravan.astro.wisc.edu/protostars/} to estimate the
contributions to the SED from the stellar photosphere, circumstellar
disk, and circumstellar envelope, and to provide insight into their
physical nature.  As well as specifying the source flux vs.
wavelength, the fitting tool also allows one to input a range of
distances and interstellar extinctions (A$_V$).  We have adopted a
distance of 500--600~pc and a possible range of interstellar A$_V$ of
0--10~mags.  The fitting process produces reasonably good fits to the
data, however, it is judicious to note that because of the many
free parameters, the model may clearly produce a good fit to the data
which is not truly representative of the physical nature of the
source.  Nevertheless, such a procedure does provide insight into the
quiescent nature of the source and hence the SED and model fits are
considered below.

The SED of \vsx\ (Figure~\ref{seds}) rises steeply from the optical to
$\sim$3~$\mu$m where it flattens out to $\sim$100~$\mu$m then declines
into the mm-wave regime.  Using the above constraint on distance and
interstellar extinction, we find that the fitting procedure produces
the best-fit parameters shown in Figure~\ref{seds} as model \#3009214
(solid line).  This model has a mass, effective temperature, and total
luminosity consistent with an intermediate-mass Herbig Ae/Be star i.e.
M$_*$=3.5~M$_{\odot}$, T$_*$=13000~K, and L$_{tot}$=126~L$_{\odot}$,
respectively.  The age of the star is {\it
  t$_*$}$\sim$4$\times$10$^{6}$~yrs and it has a total
(circumstellar~+~interstellar) extinction of A$_V$=35~mags.  However,
almost all of this extinction is unrealistically required to be of
interstellar origin.  

If we place additional constraints on the best-fit procedure,
specifically, M$_*<$2~M$_*$, the best-fit model \#3019063 (dashed
line) has M$_*$=0.7~M$_{\odot}$, T$_*$=3850~K,
L$_{tot}$=14~L$_{\odot}$, and A$_V$=12~mags.  However, the source age
for this model is small at {\it t$_*$}$\sim$2000~yrs.  If we
additionally constrain {\it t$_*$} to be between 10$^4$ and
10$^7$~yrs, the best-fit model \#3015948 (dotted line) has
M$_*$=1.9~M$_{\odot}$, T$_*$=4350~K, L$_{tot}$=30~L$_{\odot}$, and
A$_V$=25~mags.  The age of the young star in this model is
1.5$\times$10$^{5}$~yrs.  Inspection of Figure~\ref{seds} shows that
models \#3009214 and \#3019063 fit the data reasonably well. However,
model \#3015948 under-estimates the flux from 5--11~$\mu$m and
over-estimates it from 11--100~$\mu$m.  The full list of the above
model values are given in Table~\ref{sedbest}.

Taking a somewhat different approach, which is the one adopted here,
if constrain the interstellar A$_V$ component to be $<$10~mags (as
suggested by the JHK c--c diagram) and average the top 20 best-fit
models that satisfy this constraint, we obtain the final column of
values in Table~\ref{sedbest}.  As can be seen, the top 20 models show
free-parameters with, in most cases, reasonable 1$\sigma$ standard
deviations.  These values are most likely the best we can obtain for
\vsx.  The source age is again very small at
2.1$\pm$0.2$\times$10$^{3}$~yrs, however, it is calculated from the
derived values of stellar radius, temperature, and mass and so are
model dependent (B.~Whitney 2011, private communication).  All we can
likely say is that the top 20 A$_V$-constrained models indicate that
\vsx\ is relatively young.

\subsection{Accretion Luminosities and Rates}
It is possible to investigate the accretion luminosity, L$_{acc}$, and
mass accretion rate, \.M, using the V-band flux excess, the H$\alpha$
FW10\% values, and the observed Pa$\beta$ and Br$\gamma$ fluxes as we
have previously attempted for V1647~Ori (Aspin, Beck \& Reipurth 2008)
and EX~Lup (Aspin et al. 2010).  However, the value of A$_V$ to adopt
in the post-outburst period is critically important to the results
obtained.  In the JHK c-c diagram (Figure~\ref{ccs}a), the location of
\vsx\ on UT 2010 September 5 (point \#4) was close to the upper end of
the CTTS locus.  This suggests that soon after the outburst occurred,
the source was not particularly reddened by overlying extinction.  If
we deredden its location to a slightly extended CTTS locus (point B),
we find A$_V\sim$2~mag.  This is in contrast to its location in 2006
when it was both highly reddened and possessed a significant K-band
thermal excess.  On UT 2010 November 25 (point \#8), it appears that
the overlying extinction had increased with respect to that on
October~2.  Dereddening to the CTTS locus gives an A$_V\sim$7~mags.
Since the 1.282~$\mu$m Pa$\beta$ flux is significantly more affected
by extinction than the 2.166~$\mu$m Br$\gamma$ flux, we only use the
fluxes from the longer wavelength line below.

Above we noted that the observed Br$\gamma$ line fluxes were 9 and
2.3$\times$10$^{-14}$~ergs~cm$^{-2}$~s$^{-1}$ on UT 2010 October 2 and
November 26, respectively.  Dereddening these values by A$_V$=2 and
7~mags, respectively, gives fluxes of 1.1$\times$10$^{-13}$ and
4.7$\times$10$^{-14}$~ergs~cm$^{-2}$~s$^{-1}$~s$^{-1}$.  Following
Muzerolle et al. (1998) and Gullbring et al.  (1998), we can derive
L$_{acc}$ and \.M, however, for this determination we have to adopt
values for the parameters M$_*$ (the mass of the young star), L$_*$
(the stellar luminosity), T$_*$ (the star's effective temperature),
and {\it r}$_{in}$ (the inner radius of the accretion disk).  As we
have seen, the SED fitting gave ambiguous results.  However, if we
adopt the most likely physical nature for \vsx, that of a young,
low-mass T~Tauri star, we can constrain the above parameters to be
M$_*\sim$0.7~M$_{\odot}$, T$_*\sim$4000~K, R$_*\sim$5--6~R$_{\odot}$
and using the standard luminosity, temperature and stellar radius
relationship, L$_*\sim$7~L$_{\odot}$.  Gullbring et al. (1998) adopted
{\it r}$_{in}\sim$5~R$_*$ as a typical value for CTTSs and we also
adopt this value.  For the UT 2010 October 2 Br$\gamma$ flux and an
A$_V$=2, we derive L$_{acc}$=5.9$\pm$2~L$_{\odot}$ and
\.M=1.8$\pm$0.4$\times$10$^{-6}$~M$_{\odot}$~yr$^{-1}$.  For the UT
2010 November 26 Br$\gamma$ flux and an A$_V$=7, these values become
L$_{acc}$=2.0$\pm$0.7~L$_{\odot}$ and
\.M=6.3$\pm$0.5$\times$10$^{-7}$~M$_{\odot}$~yr$^{-1}$.  If we perform
the same analysis with the Br$\gamma$ fluxes from Covey et al. (2011)
we obtain L$_{acc}$ values of 2.1$\pm$0.5 and 5.2$\pm$1.5~L$_{\odot}$
for UT July 14 and July 18, respectively (using A$_V$=12~mags on both
dates).  Using the same stellar/disk parameters, these convert to \.M
rates of 6.4$\pm$1.5$\times$10$^{-7}$ and
1.6$\pm$0.5$\times$10$^{-6}$~M$_{\odot}$~yr$^{-1}$.  The above results
are shown in tabular form in Table~\ref{accret}. Although the derived
values are dependent on the assumed star/disk parameters, using the
same values for all four epochs allows us to inter-compare our
results. We find that the accretion luminosity and mass accretion
rate increases by a factor 2.5 between UT 2010 July 14 and 18,
increased still further by a factor 1.1 between UT July 18 and October
2, then declined by a factor 2.9 between UT 2010 October 2 and
November 26.  If we integrate the accretion rate over the period UT
2010 July 14 to November 26 we obtain a total mass accreted of
$\sim$2$\times$10$^{-4}$~M$_{\odot}$.

Figure~\ref{full} shows that \vsx\ has a deconvolved H$\alpha$ FW10\%
of $\sim$550~km~s$^{-1}$.  Using equation (1) of Natta et al.  (2004)
we obtain a value for \.M of
~$\sim$3$\times$10$^{-8}$~M$_{\odot}$~yr$^{-1}$.  The uncertainties on
this value (from Natta et al. equation 1) give a possible range of
rates of 6$\times$10$^{-9}$ to
1$\times$10$^{-7}$~M$_{\odot}$~yr$^{-1}$.  Even the high end of this
range is 10$\times$ smaller than the value above derived using the
dereddened Br$\gamma$ flux.

Despite the significant uncertainties of the above analysis, we can
conclude that the above estimates of \.M for \vsx\ during outburst are
considerably lower than the expected values in a FUor eruption e.g.
10$^{-5}$ to 10$^{-4}$~M$_{\odot}$~yr$^{-1}$.  We find, therefore,
that the outburst of \vsx\ seems more consistent with those observed
in EXors than in FUors.

\subsection{Indicators of Outflow and Stellar/Disk Winds}
As discussed by Covey et al. (2011), Hartigan, Edwards, \& Ghandour
(1995) presented a relationship between mass-loss rate and the
luminosity of certain forbidden optical emission lines related to
shock-excitation by outflowing gas.  Using their equations A8 and A10,
Covey et al. estimated the mass outflow rate, \.M$_{wind}$, to be
7$\times$10$^{-7}$ and 2$\times$10$^{-6}$~M$_{\odot}$~yr$^{-1}$ from
the 6716~\AA\ [\ion{S}{2}] line and the 6300~\AA\ [\ion{O}{1}] line,
respectively.  Stressing the large uncertainties in these values due
to the assumed values for visual extinction and the physical
properties of the shock region, they nonetheless showed that
significant mass outflow is occurring during periods of intense mass
accretion.  Using both the [\ion{O}{1}] line fluxes from Covey et al.
(2011) and or own from Table\ref{fluxes}, we have re-calculated
mass-loss rates using a standard set of input parameters.  The only
quantity we have varied between epochs is the overlying visual
extinction.  This we estimated from the JHK c-c diagram shown in
Figure\ref{ccs}a.  Table~\ref{massloss} shows the estimated mass
loss-rates for the four epochs for which [\ion{O}{1}] emission line
fluxes are available.  We only use the [\ion{O}{1}] line here since it
is stronger than the [\ion{S}{2}] lines present and, hence, the fluxes
are considered more reliable.  For all calculations, we assume values
for the electron density, N$_e$, equal to the [\ion{O}{1}] critical
density, N$_c$=1.97$\times$10$^{6}$~cm$^{-3}$, the projected size of
the aperture on the plane of the sky,
l$_{perp}$=2.62$\times$10$^{15}$~cm (corresponding to 0$\farcs$3 at
600~pc), and the wind velocity component in the plane of the sky,
v$_{perp}$=150~km~s$^{-1}$.  We note that \.M$_{wind}$ scales as
(1+N$_{c}$/N$_{e}$), l$_{perp}^{-1}$, and v$_{perp}^{-1}$.  For the
observed [\ion{O}{1}] fluxes and assumed values of A$_V$, we estimate
mass-loss rates of 1--2$\times$10$^{-6}$~M$_{\odot}$~yr$^{-1}$ in July
2010 dropping very significantly to
2--3$\times$10$^{-9}$~M$_{\odot}$~yr$^{-1}$ in September 2010, and
increasing to $\sim$4$\times$10$^{-8}$~M$_{\odot}$~yr$^{-1}$ in
November 2010.  Clearly, these values are highly uncertain and
evidently dependent on the assumed A$_V$ value.  However, the overall
trend indicates that the mass-loss rate in the wind was highest in
July 2010 (R$_{PTF}$=15.7), prior to the September peak in optical
brightness, then declined by three orders of magnitude at the time of
the September 2010 brightness peak (R$_{PTF}$=13.6), and then
increased by a factor $\sim$10 as the source faded back to a more
quiescent-like optical brightness in November 2010 (R$_{PTF}$=18.0).
This behavior is somewhat peculiar since the expectation would more
likely be that the mass-loss rate would increase significantly
immediately after the dramatic increase in brightness occurred.

\section{Summary} 
What we have learned about the young star \vsx\ is summarized
below.

\begin{figure}[t]
\begin{center}
\includegraphics*[angle=0,scale=0.9]{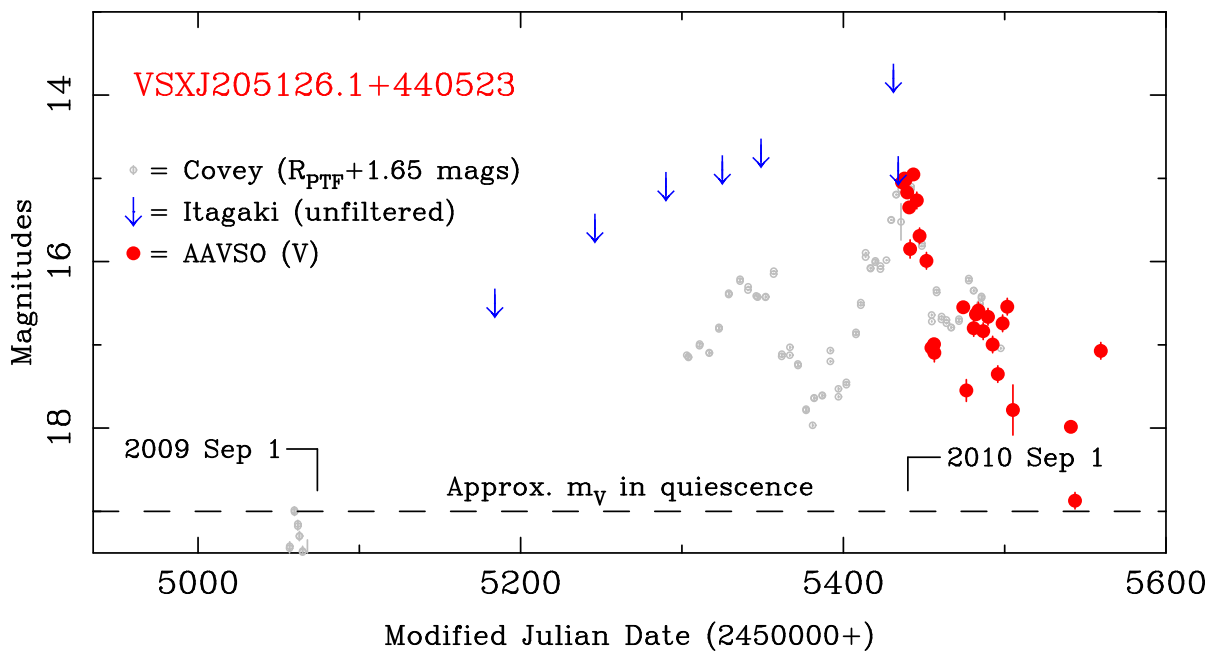} 
\caption{The optical V-band light-curve of \vsx\ 
using data from the AAVSO database.  The plot is up to date 
to the time of paper submission.  The quiescent V-band magnitude is shown 
as a horizontal dashed line. The AAVSO observations (red filled 
circles) have been
averaged to produce one composite measurement per night.  Error bars are shown
on each point and are either calculated from the spread of values throughout
each specific night or, in the case of only one observation per night, the
quoted observational uncertainty in the AAVSO database.  The downward 
facing blue arrows indicate points from Itagaki (2010) in unfiltered light and 
hence are upper-limits.  The small open grey circles are the R-band data 
from Covey et al. (2011) shifted by +1.65~mags (to approximately 
the same level as the V-band
measurements) and are shown to indicate the level of variability observed.
\label{lcs}}
\end{center}
\end{figure}

\begin{figure}[t] 
\vspace{3cm}
\begin{center}
\includegraphics*[angle=270,scale=0.9]{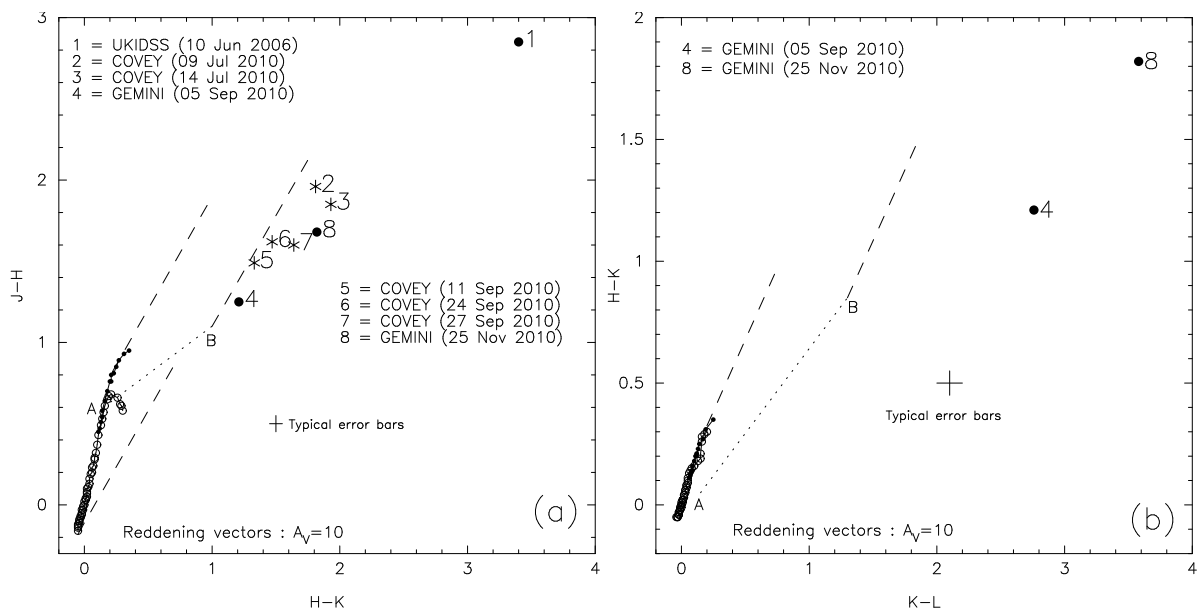} 
\caption{NIR color-color diagrams showing the location of \vsx.  
Left (a) shows the J--H vs. H--K' diagram and points \#1 and \#1 are
\vsx\ from two epochs.  On the right (b) is a H--K' vs. K'--L' plot 
showing \vsx\ from one epoch (\#2).  Typical photometric
uncertainties are shown and are of the order $\pm$0.1 mags.  The
location of zero-age main sequence dwarfs and giants are shown in
both plots.  Reddening vectors extend from
the extremes of the dwarf and giant loci with a length A$_V$=10 mags. 
The extinction law is from the Robitaille et al. (2007) SED models.
The dashed lines show the Greene \& Meyer (1995) loci for classical
and weak-line T~Tauri stars.  Extending from the end of these loci 
is an additional reddening vector (dot-dashed line).  
\label{ccs}}
\end{center}
\end{figure}

\begin{figure}[t] 
\vspace{3cm}
\begin{center}
\includegraphics*[angle=0,scale=0.7]{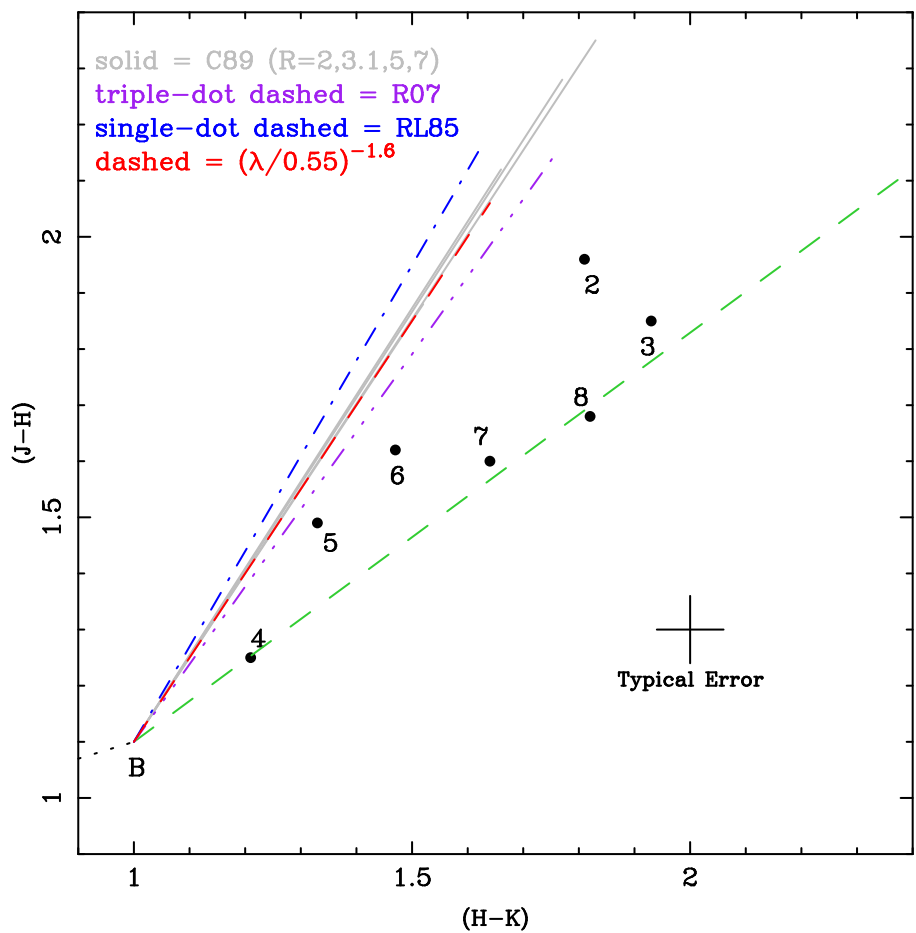} 
\caption{NIR JHK color-color diagram showing the progression of J--H 
and H--K color changes from point \#1 to \#8.  Point \#1 is outside the
plotted area and along the dashed (green) line.  Also shown are a 
selection of reddening vectors representing different reddening laws.
The grey lines show the Cardelli, Clayton, \& Mathis (1989) reddening
law for four different values of R, the ratio of total to selective 
extinction.  The dashed (red) line shows a generic reddening law 
with an exponent of --1.6.  The single-dot dashed (blue) line is the
Rieke \& Lebofsky (1985) reddening law.  Finally, the triple-dot dashed 
(purple) line shows the reddening law adopted by Robitaille et al. (2007)
in their SED models.  Although there is significant divergence of these
reddening vectors at large A$_V$, none alone can explain the changes 
observed in the \vsx\ observations. 
\label{ccs2}}
\end{center}
\end{figure}

\begin{figure}[t] 
\vspace{3cm}
\begin{center}
\includegraphics*[angle=270,scale=1.0]{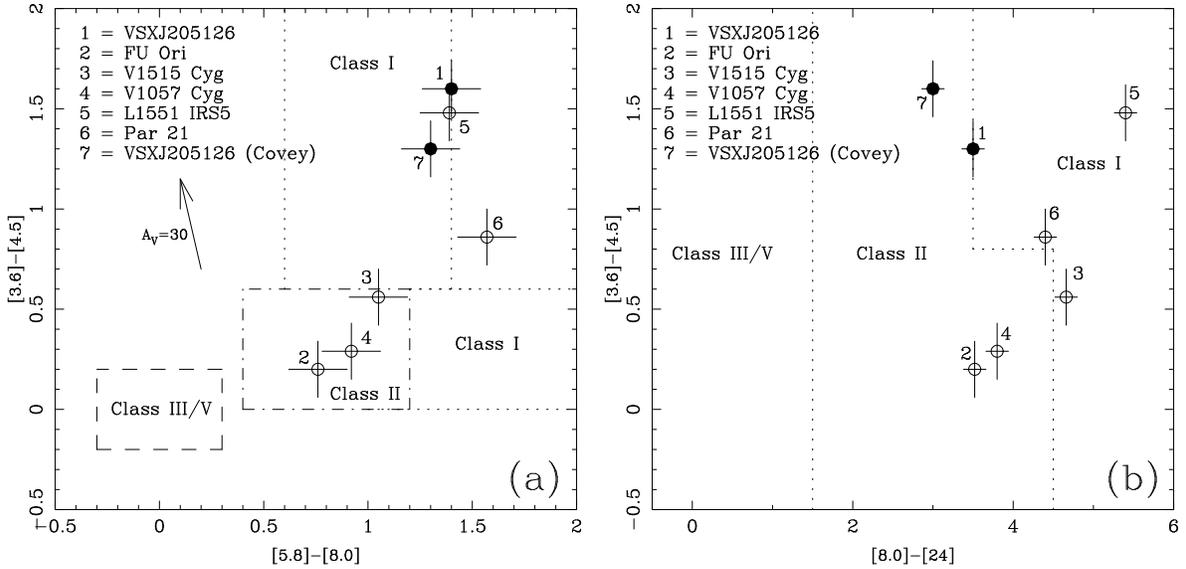} 
\caption{Spitzer IRAC and IRAC/MIPS color-color (c-c) diagrams showing the 
location of  \vsx.  Left (a) shows the [3.6]--[4.5] diagram and point \#2 
(filled circles) is \vsx.  Points \#2--6 (open circles) are the 
FUors/FUor-like objects FU~Ori, V1515~Cyg, V1057~Cyg, L1551~IRS5, and 
Par~21, respectively.  On the right (b) is a [3.6]-[4.5] vs. [8.0]--[24] plot 
with the same source identification numbers. Typical photometric 
uncertainties are shown and are of the order $\pm$0.1 mags.
The evolutionary class boundaries from Allen et al. (2002) are shown as 
are reddening vectors for A$_V$=30 mags. \vsx\ is located well 
inside the Class~I regime in the IRAC c-c diagram. In the IRAC/MIPS plot 
\vsx\ is on the Class~I--II boundary. 
\label{spitccs}}
\end{center}
\end{figure}

\begin{figure}[t] 
\begin{center}
\includegraphics*[angle=270,scale=1.3]{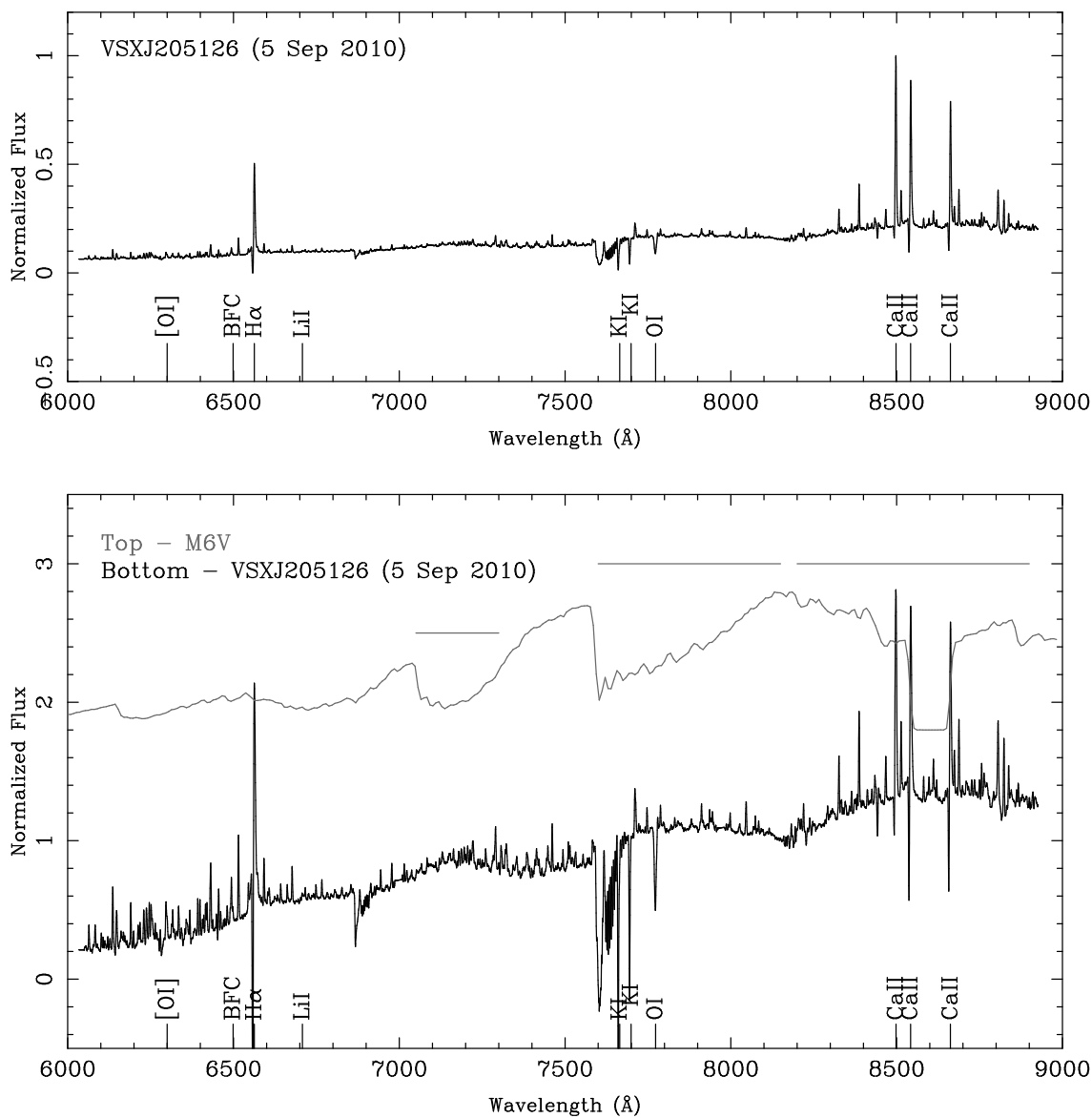} 
\caption{Top: The Gemini-North GMOS spectra of \vsx\ covering the 
wavelength range 6000--9000~\AA.  The source shows H$\alpha$ and 
far-red \ion{Ca}{2} triplet emission, both with significantly 
blue-shifted absorption components. Both \ion{K}{1} and \ion{O}{1} 
are seen in absorption and many other weaker lines are in emission.  
BFC refers to the blended \ion{Ba}{2}, \ion{Fe}{1}, and \ion{Ca}{1} 
lines at $\sim$6500~\AA. Bottom: The same spectrum of \vsx\ but
displayed on a log$_{10}$ scale.  Also shown is a spectrum of an M6~V
star to show the location of the TiO absorption bands which are in
emission in \vsx.
\label{full}}
\end{center}
\end{figure}

\begin{figure}[t] 
\begin{center}
\includegraphics*[angle=270,scale=1.3]{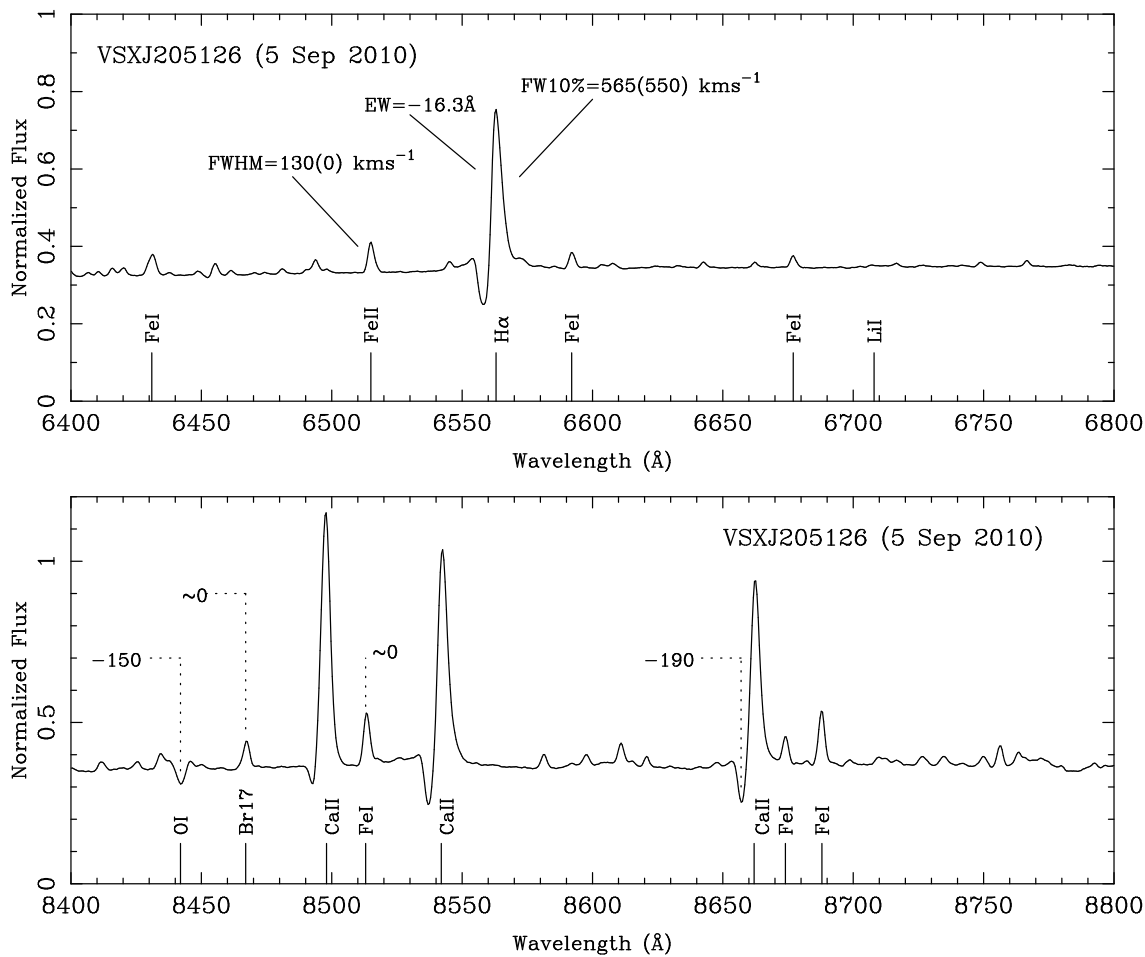} 
\caption{Top, the 6400--6800~\AA\ region of the spectra shown in
Figure~\ref{full}.  H$\alpha$ is  strongly in emission 
and clearly shows an associated blue-shifted absorption component 
creating a P~Cygni profile.  The velocity of the minimum absorption is 
at a blue-shifted velocity of --216~km~s$^{-1}$.  The \ion{Ba}{2},
\ion{Fe}{1}, \ion{Ca}{1} blend at $\sim$6500~\AA\ is not present, however,   
emission lines from \ion{Fe}{1} and \ion{Fe}{2} are seen.  Li~I absorption 
is not present and must be heavily veiled by emission.  Bottom, the 
8400--8800~\AA\ region of the spectra shown in Figure~\ref{full}.  This 
region contains the far-red \ion{Ca}{2} triplet lines.  \vsx\ shows 
the far-red \ion{Ca}{2} triplet lines strongly in emission with 
blue-shifted absorption components with the center of the 
absorption occurring at --190~km~s$^{-1}$.
\label{vhacaii}}
\end{center}
\end{figure}

\begin{figure}[t] 
\begin{center}
\includegraphics*[angle=270,scale=1.3]{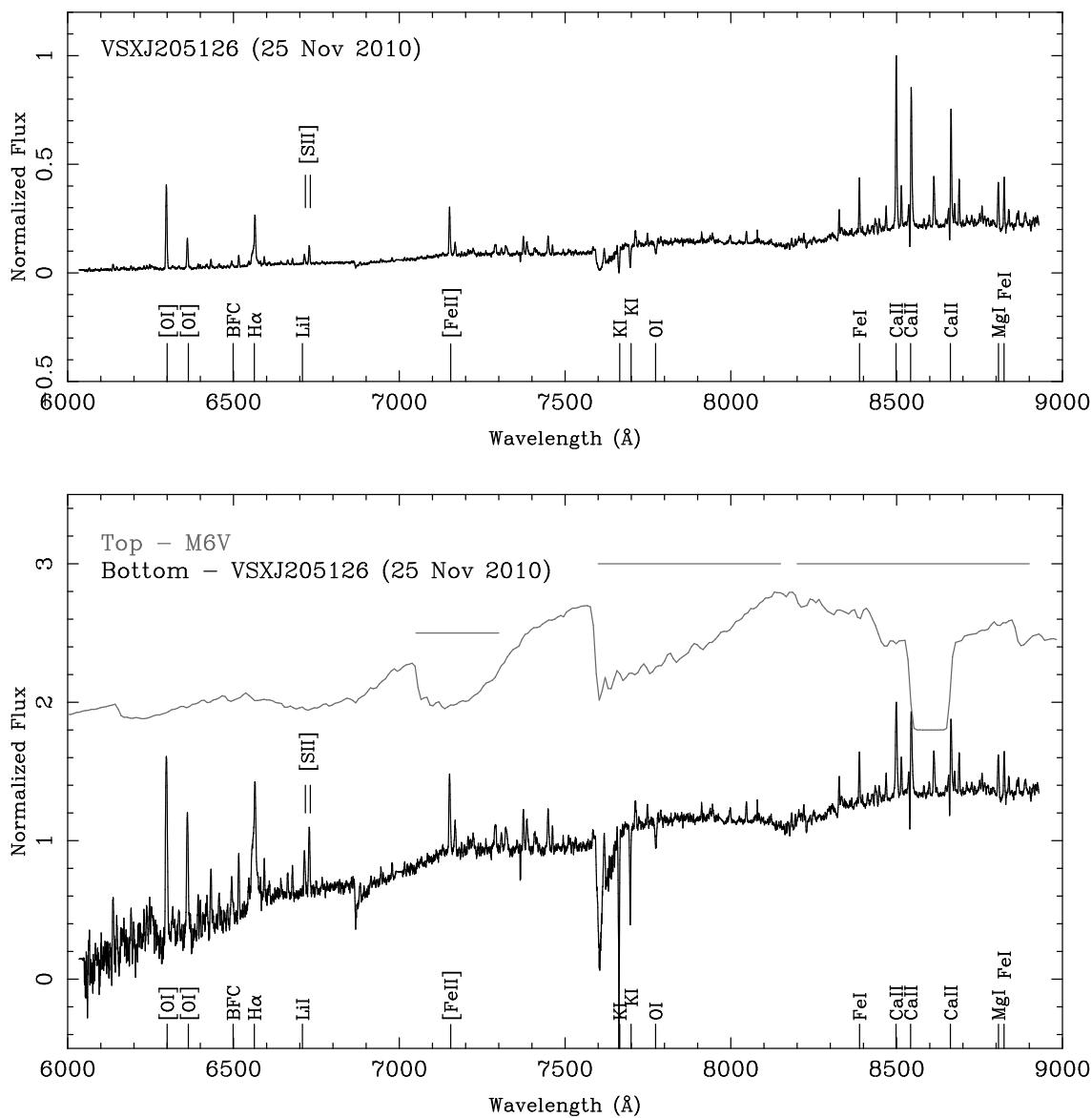} 
\caption{Top: The Gemini-North GMOS spectra of \vsx\ from UT 
2010 November 25 covering the wavelength range 6000--9000~\AA.  
The source shows H$\alpha$ and far-red \ion{Ca}{2} triplet emission 
with \ion{K}{1} and \ion{O}{1} in absorption. 
BFC refers to the blended \ion{Ba}{2}, \ion{Fe}{1}, and \ion{Ca}{1} 
lines at $\sim$6500~\AA. Bottom: The same spectrum of \vsx\ but
displayed on a log$_{10}$ scale.  Also shown is a spectrum of an M6~V
star to show the location of the TiO absorption bands which are in
emission in \vsx.
\label{full2}}
\end{center}
\end{figure}

\begin{enumerate}

\item The outburst of \vsx\ appears to be more representative of those
  occurring in EXor rather than FUor variables.  Its optical and NIR
  emission line spectra and shorter timescale brightness variations
  are much more like those found in EX~Lup and VY~Tau than in any of
  the classical FUors or FUor-like objects.

\item Based on {\it Spitzer} c-c diagrams, \vsx\ is designated a
  Class~I protostar consistent with the evolutionary state determined
  by Covey et al. (2011).

\item Although it is not clear when precisely the outburst in \vsx\
  began, we can obtain a timeline, albeit incomplete, of events by
  utilizing the light-curve of Covey et al. (2011) together with our
  Figure~\ref{lcs}.  In quiescence, \vsx\ appears to be optically
  faint with an R-band magnitude of 18--19.  The source had such a
  brightness in the USNO-B1 survey from 1979.9 and in September and
  October 2009 (Covey et al. 2011).  However, rather than being
  relatively constant in optical brightness in quiescence, some
  small-scale variability seems to be present (hence the above
  quiescent magnitude range).  For example, Itagaki (2010) presented
  photometry from UT 2009~May~7 which showed \vsx\ had an unfiltered
  optical magnitude of $\sim$15.4. This corresponds to R$\sim$15.8
  (see Figure~\ref{lcs} for a comparison of unfiltered to R-band
  photometry).  At some time between UT~2009~October~25 (the last 2009
  point from Covey et al. (2011) and UT~2009~December~19 (the first of
  the Itagaki 2010 points shown in Figure~\ref{lcs}) the current
  activity period began.

\item A significant fraction of the total change in optical and NIR
  brightness is likely due to the disappearance and re-formation of
  dust along our line-of-sight.  Although the changes in NIR colors do
  not correspond well to those expected from previously used
  extinction laws, the more or less linear (in log$_{10}$--space)
  change in location of \vsx\ in a JHK c-c diagram strongly suggests
  that extinction variations are the dominant factor.  Dust
  sublimation and re-formation have been previously observed in the
  eruptive variable V1647~Ori; the dust is sublimated by an accretion
  burst and then re-forms as the accretion process returns to a more
  stable state.

\item During the outburst, \vsx\ developed a strong stellar/disk
  wind seen as blue-shifted absorption on several strong emission
  lines.  For example, on UT~2010~September~5 both H$\alpha$ and the
  far-red \ion{Ca}{2} triplet lines showed velocities of
  $\sim$--200~km~s$^{-1}$ in the sub-continuum blue-shifted absorption
  minimum.  Between that date and UT 2010 November 26, the velocity of
  the minimum absorption trough on two of the \ion{Ca}{2} triplet
  lines reduced to $\sim$125~km~s$^{-1}$ (this trough disappeared on
  the other \ion{Ca}{2} line) while the sub-continuum absorption in
  H$\alpha$ became a low-velocity absorption on the blue-wing of the
  emission profile.  An emission line typically associated with
  mass-loss is the 6300~\AA\ [\ion{O}{1}] line.  The dereddened
  [\ion{O}{1}] line fluxes suggest that the star/disk wind was
  strongest prior to the September 2010 brightness peak.  At the time
  of the September 2010 maximum, it had declined by three orders of
  magnitude but then increased in strength by a factor 10 some
  3~months later.

\item The NIR CO overtone bandhead emission from UT 2010 October 2
  indicates that the emitting region has a temperature,
  T$_{CO}\sim$3200~K, is somewhat optically thick, and has a Gaussian
  velocity dispersion, v$_g\sim$55~km~s$^{-1}$.  Approximately two
  months later, both T$_{CO}$ and v$_g$ had increased to $\sim$4200~K
  and 85~km~s$^{-1}$, respectively.  This is consistent with the
  buildup of gaseous, circumstellar material closer to the star as the
  eruption subsides.  The lack of evidence for this gaseous material
  being in the form of a circumstellar disk either suggests that the
  emitting region is located elsewhere, perhaps in accretion funnels,
  or that the disk is viewed either relatively face-on or partly
  obscured from directly view.  For a 0.7~M$_{\odot}$ central star
  with a Keplerian disk, velocities of 55 and 85~km~s$^{-1}$
  correspond orbital radii of 0.2 and 0.09~AU, respectively.

\item Optical (e.g. [\ion{S}{2}]) and NIR (e.g. H$_2$) shock-excited
  emission lines were present in the earliest Covey et al. (2011)
  spectrum from UT 2010 July 14, absent (or very weak) in our UT 2010
  September 5 spectrum, and present again in our most recent spectrum
  from UT 2010 November 25.  There is no evidence as to the origin of
  these lines but possibilities include shocks associated with the
  accretion flow or those created by outflowing gas such as in a
  collimated Herbig-Haro flow.  We cannot distinguish between these
  mechanisms at this time.

\item The pre-outburst SED of \vsx\ has been fitted using the model
  SEDs of Robitaille et al. (2006, 2007).  The best fits indicate that
  the star is at a very young evolutionary stage and is of low-mass
  with M$_{*}\sim$0.5~M$_{\odot}$, T$_*\sim$3600~K,
  L$_{tot}\sim$20~L$_{\odot}$, and R$_{*}\sim$7~R$_{\odot}$.  Due to
  its age, the star still possesses a significant circumstellar
  envelope (M$_{env}\sim$0.02~M$_{\odot}$) and disk
  (M$_{disk}\sim$0.01~M$_{\odot}$).  All indications are that it is
  still in the Class~I protostar phase.

\item From an analysis of the Br$\gamma$ line fluxes from Covey et al.
  (2011) and from our spectra, we find that the accretion luminosity
  and rate fall in the ranges L$_{acc}$=1--6~L$_{\odot}$ and
  \.M=0.4--6$\times$10$^{-6}$~M$_{\odot}$~yr$^{-1}$, respectively.
  Although these values depend on the assumed A$_V$ at the time of
  observations, we can say with reasonable confidence that these
  accretion parameter values are significantly lower than found in
  FUors/FUor-like objects and more like those occurring in EXor
  outbursts.

\end{enumerate}

It remains to be seen how \vsx\ evolves and whether it continues to
exhibit short-timescale variability or returns to a more quiescent
state.  Nevertheless, our conclusion is that it appears to be another
example of an EXor variable similar in nature to EX~Lup although
apparently significantly younger.  Continuous optical monitoring of
the star by the AAVSO observers will clearly greatly assist in a
better understanding of this enigmatic young stellar object.

\vspace{0.3cm}

{\bf Acknowledgments} We are extremely grateful to the referee,
Barbara Whitney, for many useful comments on the manuscript, in
particular, on the interpretation of the best-fit SED models.  We wish
to express our sincere thanks to Nancy Levinson for allocating the
Director's time used to obtain our GMOS-N, NIRI, and NIFS Gemini-North
observations.  We are grateful to the AAVSO and their dedicated
observers who continue to provide high quality optical photometry of
\vsx.  We are pleased to acknowledge Kevin Covey for valuable
discussions and providing some additional information on his recent
paper and on the source itself.  We thanks Bo Reipurth for useful
discussions and critical reading of the manuscript.  This project was
supported by the Gemini Observatory, which is operated by the
Association of Universities for Research in Astronomy, Inc., on behalf
of the international Gemini partnership of Argentina, Australia,
Brazil, Canada, Chile, the UK, and the US.  This work is based in part
on data obtained as part of the UKIRT Infrared Deep Sky Survey.  This
publication makes use of data products from the Two Micron All Sky
Survey, which is a joint project of the University of Massachusetts
and the Infrared Processing and Analysis Center/California Institute
of Technology, funded by the National Aeronautics and Space
Administration and the National Science Foundation.  This work is
based in part on observations made with the Spitzer Space Telescope,
which is operated by the Jet Propulsion Laboratory, California
Institute of Technology under a contract with NASA. Support for this
work was provided by NASA through an award issued by JPL/Caltech.
This research is based in part on observations with AKARI, a JAXA
project with the participation of ESA.



\clearpage

\begin{deluxetable}{lclr}
\tablecaption{Observation Log\label{obslog}}
\tablewidth{0pt}
\tablehead{
\colhead{UT Date} & 
\colhead{JD} &
\colhead{Telescope/} & 
\colhead{Details} \\
\colhead{} &
\colhead{} &
\colhead{Instrument} &
\colhead{}}
\startdata
2010 Sep  5 & 2455445 & Gemini-N/GMOS & r' imaging \\
2010 Sep  5 & 2455445 & Gemini-N/GMOS & B600 spectroscopy \\
2010 Sep  5 & 2455445 & Gemini-N/NIRI & J,H,K',L' photometry \\
2010 Oct  2 & 2455472 & Gemini-N/NIFS & K-band spectroscopy \\
2010 Oct  3 & 2455473 & Gemini-N/NIFS & J,H spectroscopy \\
2010 Nov 25 & 2455525 & Gemini-N/GMOS & B600 spectroscopy \\
2010 Nov 25 & 2455525 & Gemini-N/NIRI & J,H,K',L' photometry \\
2010 Nov 26 & 2455526 & Gemini-N/NIFS & K-band spectroscopy \\
2010 Nov 27 & 2455527 & Gemini-N/NIFS & H-band spectroscopy \\
\enddata

\end{deluxetable}

\clearpage

\begin{deluxetable}{lcccr}
\tablecaption{NIR Photometry\label{nirphot}}
\tablewidth{0pt}
\tablehead{
\colhead{Filter} & 
\colhead{2000/06/10} &
\colhead{2006/06/10} &
\colhead{2010/09/05} &
\colhead{2010/11/25} \\
\colhead{} &
\colhead{(2MASS)} &
\colhead{(UKIDSS)} &
\colhead{(Gemini)} &
\colhead{(Gemini)}}
\startdata
J     & $>$15.8 & 19.40$\pm$0.09 & 10.58$\pm$0.1  & 13.13$\pm$0.1  \\
H     & $>$15.1 & 16.55$\pm$0.02 &  9.33$\pm$0.1  & 11.45$\pm$0.1  \\
K'    & $>$14.3 & 13.15$\pm$0.01 &  8.12$\pm$0.1  &  9.63$\pm$0.1  \\
L/L'  & --      & --             &  5.36$\pm$0.1  &  6.05$\pm$0.1  \\
J-H   & --      & 2.85$\pm$0.09  &  1.25$\pm$0.14 &  1.68$\pm$0.14 \\
H-K'  & --      & 3.40$\pm$0.02  &  1.21$\pm$0.14 &  1.82$\pm$0.14 \\
K'-L' & --      & --             &  2.76$\pm$0.14 &  3.58$\pm$0.14 \\
\enddata

\end{deluxetable}

\clearpage

\begin{deluxetable}{lc|ccc|ccc|}
\rotate
\tablecaption{Optical Spectral Features with EW~$\ge$~1~\AA\ on One of the Two Observation Dates\label{fluxes}}
\tablewidth{0pt}
\tablehead{
\colhead{Line} & 
\colhead{$\lambda$} & 
\colhead{EW\tablenotemark{a}} &
\colhead{Line Flux} &
\colhead{Continuum} &
\colhead{EW\tablenotemark{a}} &
\colhead{Line Flux} &
\colhead{Continuum} \\
\colhead{} &
\colhead{(\AA)} &
\colhead{(\AA)} &
\colhead{(ergs~cm$^{-2}$~s$^{-1}$)} &
\colhead{(ergs~cm$^{-2}$~s$^{-1}$~\AA$^{-1}$)} &
\colhead{(\AA)} &
\colhead{(ergs~cm$^{-2}$~s$^{-1}$)} &
\colhead{(ergs~cm$^{-2}$~s$^{-1}$~\AA$^{-1}$)} \\
\colhead{} & 
\colhead{} & 
\multicolumn{3}{c}{05 September 2010} & 
\multicolumn{3}{c}{25 November 2010}}
\startdata
                  \ion{Fe}{1}  & 6136 &    --1.5 &   1.53(--14)\tablenotemark{b} &   1.02(--14) &    --2.4 &   2.64(--16) &   1.10(--16) \\
                  \ion{Fe}{2}  & 6148 &    --1.0 &   1.03(--14) &   1.03(--14) &    --0.9 &   1.01(--16) &   1.12(--16) \\
                  \ion{Fe}{1}  & 6192 &    --0.8 &   8.53(--15) &   1.07(--14) &    --1.0 &   1.19(--16) &   1.19(--16) \\
                  \ion{Fe}{2}  & 6238 &    --0.9 &   9.84(--15) &   1.09(--14) &    --1.6 &   1.99(--16) &   1.24(--16) \\
                  \ion{Fe}{2}  & 6247 &    --0.8 &   8.78(--15) &   1.10(--14) &    --1.1 &   1.37(--16) &   1.25(--16) \\
             $[$\ion{O}{1}$]$  & 6300 &    --1.3 &   1.44(--14) &   1.11(--14) &   --38.9 &   4.94(--15) &   1.27(--16) \\
             $[$\ion{O}{1}$]$  & 6364 &    --0.6 &   6.77(--15) &   1.13(--14) &   --13.3 &   1.72(--15) &   1.30(--16) \\
                  \ion{Fe}{1}  & 6393 &    --0.7 &   8.02(--15) &   1.15(--14) &    --1.7 &   2.24(--16) &   1.32(--16) \\
                  \ion{Fe}{1}  & 6399 &    --0.7 &   8.05(--15) &   1.15(--14) &    --1.5 &   1.99(--16) &   1.33(--16) \\
                  \ion{Fe}{2}  & 6433 &    --2.3 &   2.73(--14) &   1.19(--14) &    --3.1 &   4.27(--16) &   1.38(--16) \\
      \ion{Ca}{2}/\ion{Fe}{2}\tablenotemark{c}  & 6456 &    --0.8 &   9.74(--15) &   1.22(--14) &    --1.5 &   2.14(--16) &   1.43(--16) \\
                  \ion{Fe}{1}  & 6496 &    --0.9 &   1.16(--14) &   1.28(--14) &    --2.4 &   3.66(--16) &   1.52(--16) \\
                  \ion{Fe}{2}  & 6516 &    --2.6 &   3.43(--14) &   1.32(--14) &    --3.6 &   5.68(--16) &   1.58(--16) \\
             $[$\ion{N}{2}$]$  & 6548 &    --0.7 &   9.61(--15) &   1.37(--14) &    --1.3 &   2.16(--16) &   1.66(--16) \\
                    H$\alpha$  & 6563 &   --16.3 &   2.28(--13) &   1.40(--14) &   --27.6 &   4.68(--15) &   1.70(--16) \\
             $[$\ion{N}{2}$]$  & 6592 &    --1.2 &   1.72(--14) &   1.44(--14) &    --2.0 &   3.52(--16) &   1.76(--16) \\
                  \ion{Fe}{1}  & 6664 &      ...\tablenotemark{d} &          ... &          ... &    --1.0 &   1.84(--16) &   1.84(--16) \\
                  \ion{Fe}{1}  & 6677 &    --0.8 &   1.19(--14) &   1.48(--14) &    --1.3 &   2.39(--16) &   1.84(--16) \\
             $[$\ion{S}{2}$]$  & 6716 &      ... &          ... &          ... &    --2.8 &   5.15(--16) &   1.84(--16) \\
             $[$\ion{S}{2}$]$  & 6731 &      ... &          ... &          ... &    --5.1 &   9.37(--16) &   1.84(--16) \\
            $[$\ion{Fe}{2}$]$  & 7155 &      ... &          ... &          ... &    --8.4 &   2.40(--15) &   2.86(--16) \\
                  \ion{Fe}{2}  & 7169 &      ... &          ... &          ... &    --2.2 &   6.39(--16) &   2.90(--16) \\
            $[$\ion{Ca}{2}$]$  & 7291 &    --1.5 &   2.90(--14) &   1.93(--14) &    --2.2 &   6.75(--16) &   3.07(--16) \\
                  \ion{Fe}{2}  & 7308 &    --0.6 &   1.15(--14) &   1.91(--14) &    --1.0 &   3.06(--16) &   3.06(--16) \\
            $[$\ion{Ca}{2}$]$  & 7324 &    --0.9 &   1.70(--14) &   1.89(--14) &    --2.2 &   6.73(--16) &   3.06(--16) \\
                 \ion{Na}{1}?\tablenotemark{e}  & 7374 &      ... &          ... &          ... &    --2.8 &   8.47(--16) &   3.03(--16) \\
            $[$\ion{Fe}{2}$]$  & 7388 &    --0.8 &   1.47(--14) &   1.84(--14) &    --2.3 &   6.94(--16) &   3.02(--16) \\
                 \ion{Si}{1}?\tablenotemark{e}  & 7409 &      ... &          ... &          ... &    --1.3 &   3.91(--16) &   3.01(--16) \\
                  \ion{Fe}{2}  & 7448 &    --0.6 &   1.11(--14) &   1.85(--14) &    --3.6 &   1.08(--15) &   3.00(--16) \\
                  \ion{Fe}{2}  & 7461 &    --1.2 &   2.23(--14) &   1.86(--14) &    --1.2 &   3.61(--16) &   3.01(--16) \\
                  \ion{Fe}{1}  & 7618 &      ... &          ... &          ... &    --4.1 &   1.41(--15) &   3.43(--16) \\
                   \ion{K}{1}  & 7665 &     +3.8 &   8.58(--14) &   2.26(--14) &     +4.3 &   1.59(--15) &   3.70(--16) \\
                   \ion{K}{1}  & 7699 &     +2.7 &   6.33(--14) &   2.34(--14) &     +3.2 &   1.25(--15) &   3.91(--16) \\
                  \ion{Fe}{2}  & 7711 &    --1.9 &   4.51(--14) &   2.37(--14) &    --2.4 &   9.56(--16) &   3.98(--16) \\
                   \ion{O}{1}\tablenotemark{f}  & 7774 &     +3.3 &   8.26(--14) &   2.50(--14) &     +1.9 &   8.17(--16) &   4.30(--16) \\
                  \ion{Fe}{1}  & 7998 &      ... &          ... &          ... &    --1.1 &   4.83(--16) &   4.39(--16) \\
                  \ion{Fe}{1}  & 8046 &    --0.9 &   2.20(--14) &   2.45(--14) &    --1.2 &   5.13(--16) &   4.28(--16) \\
                  \ion{Fe}{1}  & 8219 &    --0.6 &   1.50(--14) &   2.50(--14) &    --1.1 &   4.93(--16) &   4.48(--16) \\
                  \ion{Fe}{1}  & 8327 &    --1.4 &   3.97(--14) &   2.84(--14) &    --2.9 &   1.52(--15) &   5.23(--16) \\
           \ion{H}{1} (Pa~24)  & 8334 &      ... &          ... &          ... &    --1.1 &   5.81(--16) &   5.28(--16) \\
                  \ion{Fe}{1}  & 8388 &    --2.7 &   8.19(--14) &   3.03(--14) &    --4.3 &   2.43(--15) &   5.66(--16) \\
           \ion{H}{1} (Pa~18)  & 8435 &    --0.8 &   2.53(--14) &   3.16(--14) &    --1.5 &   8.93(--16) &   5.96(--16) \\
                   \ion{O}{1}  & 8446 &     +1.0 &   3.19(--14) &   3.19(--14) &      ... &          ... &          ... \\
                  \ion{Fe}{2}  & 8448 &      ... &          ... &          ... &    --1.2 &   7.24(--16) &   6.03(--16) \\
           \ion{H}{1} (Pa~17)  & 8467 &    --1.1 &   3.56(--14) &   3.24(--14) &    --1.8 &   1.11(--15) &   6.14(--16) \\
                  \ion{Ca}{2}  & 8498 &   --15.4 &   5.08(--13) &   3.30(--14) &   --15.6 &   9.83(--15) &   6.30(--16) \\
                  \ion{Fe}{1}  & 8514 &    --2.1 &   6.98(--14) &   3.32(--14) &    --2.9 &   1.85(--15) &   6.38(--16) \\
                  \ion{Ca}{2}  & 8543 &   --13.6 &   4.57(--13) &   3.36(--14) &   --12.6 &   8.20(--15) &   6.51(--16) \\
                  \ion{Fe}{1}  & 8611 &    --0.8 &   2.73(--14) &   3.41(--14) &    --4.3 &   2.91(--15) &   6.76(--16) \\
      \ion{Ca}{2}/\ion{Fe}{1}\tablenotemark{c}  & 8662 &   --10.5 &   3.58(--13) &   3.41(--14) &    --9.6 &   6.59(--15) &   6.87(--16) \\
                  \ion{Fe}{1}  & 8674 &    --1.2 &   4.08(--14) &   3.40(--14) &    --1.4 &   9.63(--16) &   6.88(--16) \\
                  \ion{Fe}{1}  & 8689 &    --2.1 &   7.13(--14) &   3.39(--14) &    --2.8 &   1.93(--15) &   6.89(--16) \\
                  \ion{Mg}{1}  & 8807 &    --3.6 &   1.18(--13) &   3.27(--14) &    --3.6 &   2.42(--15) &   6.73(--16) \\
                  \ion{Fe}{1}  & 8824 &    --2.1 &   6.82(--14) &   3.25(--14) &    --3.4 &   2.28(--15) &   6.72(--16) \\
                  \ion{Fe}{1}  & 8838 &    --1.1 &   3.55(--14) &   3.23(--14) &    --1.2 &   8.05(--16) &   6.71(--16) \\

\enddata
\tablenotetext{a}{Equivalent widths have associated uncertainties of 
$\pm$0.2~\AA.}
\tablenotetext{b}{Number in parentheses is exponent e.g. 1.53(--14) 
means 1.53$\times$10$^{-14}$.}
\tablenotetext{c}{Indicates blended line.}
\tablenotetext{d}{... means line not present.}
\tablenotetext{e}{Question mark indicates uncertain identification with 
strongest/most likely line listed.}
\tablenotetext{f}{Unresolved triplet.}
\end{deluxetable}

\clearpage

\begin{deluxetable}{lc|ccc|ccc|}
\rotate
\tablecaption{NIR Spectral Features with EW~$\ge$~1~\AA\ on One of the Two Observation Dates\label{fluxes2}}
\tablewidth{0pt}
\tablehead{
\colhead{Line} & 
\colhead{$\lambda$} & 
\colhead{EW\tablenotemark{a}} &
\colhead{Line Flux} &
\colhead{Continuum} &
\colhead{EW\tablenotemark{a}} &
\colhead{Line Flux} &
\colhead{Continuum} \\
\colhead{} &
\colhead{(\AA)} &
\colhead{(\AA)} &
\colhead{(ergs~cm$^{-2}$~s$^{-1}$)} &
\colhead{(ergs~cm$^{-2}$~s$^{-1}$~\AA$^{-1}$)} &
\colhead{(\AA)} &
\colhead{(ergs~cm$^{-2}$~s$^{-1}$)} &
\colhead{(ergs~cm$^{-2}$~s$^{-1}$~\AA$^{-1}$)} \\
\colhead{} & 
\colhead{} & 
\multicolumn{3}{c}{02 October 2010} & 
\multicolumn{3}{c}{26 November 2010}}
\startdata
\ion{Fe}{1} & 1.1599 & --2.0 & 1.6(--14)\tablenotemark{b} & 8.0(--15) & na\tablenotemark{c} & na & na \\
\ion{Fe}{1} & 1.1613 & --3.2 & 2.5(--14) & 8.1(--15) & na & na & na \\
\ion{Fe}{1} & 1.1644 & --2.4 & 2.0(--14) & 8.1(--15) & na & na & na \\
\ion{K}{1}  & 1.1695 & --2.7 & 2.1(--14) & 8.0(--15) & na & na & na \\
\ion{C}{1}?? & 1.1758 & --1.1 & 9.3(--15) & 8.1(--15) & na & na & na \\
\ion{Fe}{1} & 1.1788 & --2.3 & 1.9(--14) & 8.3(--15) & na & na & na \\
\ion{Mg}{1} & 1.1832 & --3.3 & 2.7(--14) & 8.3(--15) & na & na & na \\
\ion{Fe}{1} & 1.1843 & --1.2 & 1.0(--14) & 8.2(--15) & na & na & na \\
\ion{Fe}{1} & 1.1887 & --5.5 & 4.6(--14) & 8.4(--15) & na & na & na \\
?? & 1.1953 & --1.5 & 1.2(--14) & 8.3(--15) & na & na & na \\
\ion{Fe}{1} & 1.1976 & --4.3 & 3.7(--14) & 8.5(--15) & na & na & na \\
\ion{Fe}{1} & 1.1987 & --1.7 & 1.5(--14) & 8.6(--15) & na & na & na \\
\ion{Si}{1}?? & 1.1995 & --1.2 & 1.0(--14) & 8.6(--15) & na & na & na \\
\ion{Fe}{1} & 1.2035 & --2.4 & 2.0(--14) & 8.5(--15) & na & na & na \\
\ion{Fe}{1} & 1.2086 & --1.3 & 1.1(--14) & 8.3(--15) & na & na & na \\
\ion{Si}{1}?? & 1.2106 & --1.6 & 1.3(--14) & 8.3(--15) & na & na & na \\
\ion{Fe}{1} & 1.2273 & --1.6 & 1.4(--15) & 8.5(--15) & na & na & na \\
\ion{H}{1} Pa$\beta$ & 1.2822 & --11.8& 1.1(--13) & 9.2(--15) & na & na & na \\
\ion{Fe}{1} & 1.2902 & --1.0 & 9.1(--15) & 9.1(--15) & na & na & na \\
\ion{Fe}{1} & 1.3136 & --1.7 & 1.6(--14) & 9.4(--15) & na & na & na \\
\ion{Si}{1}?? & 1.3154 & --1.5 & 1.4(--14) & 9.5(--15) & na & na & na \\
\ion{Fe}{1} & 1.3292 & --1.2 & 1.2(--14) & 9.7(--15) & na & na & na \\
\ion{Fe}{1} & 1.4885 & --2.4 & 3.3(--14) & 1.4(--14) & --2.1 & 4.7(--15) & 2.2(--15) \\
\ion{Fe}{1} & 1.5031 & --4.3 & 6.0(--14) & 1.4(--14) & --3.8 & 8.6(--15) & 2.3(--15) \\
\ion{Mg}{1} & 1.5046 & --3.3 & 4.5(--14) & 1.4(--14) & --3.2 & 7.2(--15) & 2.3(--15) \\
\ion{Fe}{1} & 1.5055 & --2.6 & 3.5(--14) & 1.4(--14) & --2.7 & 6.2(--15) & 2.3(--15) \\
\ion{Fe}{1} & 1.5300 & --1.0 & 1.4(--14) & 1.4(--14) & ...\tablenotemark{d} & ... & ... \\
\ion{H}{1} Br~15 & 1.5560 & --1.1 & 1.6(--14) & 1.4(--14) & --1.2 & 3.0(--15) & 2.5(--15) \\
\ion{Fe}{1} & 1.5753 & --1.2 & 1.7(--14) & 1.5(--14) & --1.3 & 3.4(--15) & 2.6(--15) \\
\ion{Fe}{1} & 1.5771 & --2.1 & 3.1(--14) & 1.5(--14) & --1.9 & 5.0(--15) & 2.7(--15) \\ 
\ion{H}{1} Br~14 & 1.5883 & --1.2 & 1.7(--14) & 1.5(--14) & ... & ... & ... \\
\ion{Si}{1} & 1.5892 & --2.4 & 3.6(--14) & 1.5(--14) & --2.1 & 5.7(--15) & 2.7(--15) \\
\ion{Fe}{1} & 1.5963 & --1.1 & 1.6(--14) & 1.5(--14) & ... & ... & ... \\
\ion{H}{1} Br~13 & 1.6111 & --1.1 & 1.7(--14) & 1.5(--14) & ... & ... & ... \\
\ion{H}{1} Br~12 & 1.6409 & --1.7 & 2.6(--14) & 1.5(--14) & --1.7 & 5.1(--15) & 3.0(--15) \\
$[$\ion{Fe}{2}$]$ & 1.6432 &   ... &       ... &       ... & --1.2 & 3.6(--15) & 3.0(--15) \\
\ion{H}{1} Br~11 & 1.6808 & --1.9 & 3.0(--14) & 1.6(--14) & --1.8 & 5.6(--15) & 3.1(--15) \\
\ion{Mg}{1} & 1.7112 & --2.0 & 3.3(--14) & 1.7(--14) & --1.6 & 5.5(--15) & 3.4(--15) \\
\ion{H}{1} Br~10 & 1.7366 & --2.4 & 4.2(--14) & 1.8(--14) & --2.4 & 8.8(--15) & 3.7(--15) \\
\ion{Fe}{1} & 1.7477 & --0.6 & 1.1(--14) & 1.8(--14) & --1.2 & 4.4(--15) & 3.8(--15) \\
$[$\ion{Fe}{2}$]$ & 2.0152 & +1.8  & 3.0(--14) & 1.7(--14) & ... & ... & ... \\
\ion{Fe}{1} & 2.0336 &   ... &       ... &       ... & --1.1 & 5.4(--15) & 4.8(--15) \\
H$_2$~1-0 S(1) & 2.1214 & --0.7 & 1.3(--14) & 1.7(--14) & --3.0 & 1.5(--14) & 5.2(--15) \\
\ion{H}{1} (Br$\gamma$) & 2.1658 & --5.1 & 9.0(--14) & 1.8(--14) & --4.2 & 2.3(--14) & 5.5(--15) \\
\ion{Na}{1} & 2.2061 & --1.1 & 1.9(--14) & 1.8(--14) & --0.8 & 4.5(--15) & 5.8(--15) \\
\ion{Na}{1} & 2.2087 & --1.0 & 1.8(--14) & 1.8(--14) & --0.7 & 4.0(--15) & 5.8(--15) \\
H$_2$~1-0 S(0)\tablenotemark{e} & 2.2235 &... &       ... &       ... & --0.7 & 4.4(--15) & 5.9(--15) \\
H$_2$~2-1 S(1)\tablenotemark{e} & 2.2477 &... &       ... &       ... & --0.5 & 2.9(--15) & 6.1(--15) \\
CO v=2--0 & 2.2931\tablenotemark{f} & --15.1& 2.9(--13) & 1.9(--14) & --10.7 & 6.9(--14) & 6.4(--15) \\ 

\enddata
\tablenotetext{a}{Equivalent widths have associated uncertainties of 
$\pm$0.2~\AA.}
\tablenotetext{b}{Number in parentheses is exponent e.g. 1.53(--14) 
means 1.53$\times$10$^{-14}$.}
\tablenotetext{c}{na means data not available.}
\tablenotetext{d}{... means line not present.}
\tablenotetext{e}{Line included even though it has an EW~$<$~1~\AA\ 
since it is useful for comparative purposes.}
\tablenotetext{f}{EW measured from 2.2930--2.3000~$\mu$m.}
\end{deluxetable}

\clearpage

\begin{deluxetable}{lrrr}
\tablecaption{Pre-Outburst Photometry of \vsx\label{sedphot}}
\tablewidth{0pt}
\tablehead{
\colhead{Wavelength} & 
\colhead{Origin} & 
\colhead{Epoch} &
\colhead{Flux} \\
\colhead{($\mu$m)} &
\colhead{of Flux} &
\colhead{of Flux} &
\colhead{(mJy)}}
\startdata
1.25      & UKIDSS             & 2006.7 &   0.03$\pm$0.003 \\
1.65      & UKIDSS             & 2006.7 &   0.25$\pm$0.025 \\
2.20      & UKIDSS             & 2006.7 &    3.8$\pm$0.38 \\
3.6       & {\it Spitzer/IRAC} & 2006.6 &     74$\pm$7 \\
4.5       & {\it Spitzer/IRAC} & 2006.6 &    158$\pm$16 \\
5.6       & {\it Spitzer/IRAC} & 2006.6 &    653$\pm$65 \\
8.0       & {\it Spitzer/IRAC} & 2006.6 &   1200$\pm$120 \\
9.0       & {\it AKARI/IRC}    & 2006.6 &   1960$\pm$200 \\
18.0      & {\it AKARI/IRC}    & 2006.6 &   3875$\pm$457 \\
24.0      & {\it Spitzer/MIPS} & 2006.6 &   3402$\pm$340 \\
70.0      & {\it Spitzer/MIPS} & 2006.6 &   5880$\pm$588 \\
1100.0    & {\it Bolocam GPS}  & 2009.4 &    148$\pm$74 \\
& & & \\
\multicolumn{4}{c}{Additional Photometry} \\
8.28      & {\it MSX6C}        & 1996.6 &  1610$\pm$66 \\
12.0      & {\it IRAS}         & 1983   &  3390$\pm$330 \\
12.13     & {\it MSX6C}        & 1996.6 &  2490$\pm$137 \\
14.65     & {\it MSX6C}        & 1996.6 &  3045$\pm$189 \\
21.34     & {\it MSX6C}        & 1996.6 &  3140$\pm$201 \\
25.0      & {\it IRAS}         & 1983   &  6590$\pm$660 \\
60.0      & {\it IRAS}         & 1983   & 27890$\pm$1000 \\
100.0     & {\it IRAS}         & 1983   & 57350$\pm$1000 \\
\enddata

\end{deluxetable}

\clearpage

\begin{deluxetable}{lcrrrr}
\tablecaption{SED Fitting Best-Fit Parameters\label{sedbest}}
\tablewidth{0pt}
\tablehead{
\colhead{Parameter} & 
\colhead{Units} &
\colhead{Best-Fit\tablenotemark{a}} & 
\colhead{Best-Fit\tablenotemark{b}} &
\colhead{Best-Fit\tablenotemark{c}} &
\colhead{Best-Fit\tablenotemark{d}} \\
\colhead{Name} &
\colhead{} &
\colhead{Value} &
\colhead{Value} &
\colhead{Value} &
\colhead{Value}}
\startdata
Model \#                                 & & 3009214            & 3019063            & 3015948 & Top~20 \\
$\chi^{2}$                               & & 274                & 602                & 1519    & 795$\pm$14 \\
Mass & (M$_{\odot}$)                       & 3.5                & 0.7                & 1.9     & 0.52$\pm$0.43 \\
Temperature & (K)                          & 13000              & 3800               & 4400    & 3580$\pm$85 \\
Age & (yrs)                                & 4$\times$10$^{6}$  & 1.6$\times$10$^{3}$& 1.5$\times$10$^{5}$ & 2.1$\pm$0.2$\times$10$^{3}$ \\
R$_*$\tablenotemark{e} & (R$_{\odot}$)     & 2                  & 9                  & 10      & 6.9$\pm$0.3 \\
L$_{tot}$\tablenotemark{f} & (L$_{\odot}$) & 126                & 14                 & 30      & 17$\pm$2 \\
Distance & (pc)                            & 500                & 500                & 500     & 521$\pm$8 \\
Inclination\tablenotemark{g} & (degs)      & 31.8               & 31.8               & 31.8    & 50$\pm$3 \\
\.M$_{env}$\tablenotemark{h} & (M$_{\odot}$~yr$^{-1}$)      & 0                  & 4$\times$10$^{-6}$ & 3$\times$10$^{-5}$ & 3.6$\pm$0.3$\times$10$^{-6}$ \\
\.M$_{disk}$\tablenotemark{i} & (M$_{\odot}$~yr$^{-1}$)     & 8$\times$10$^{-8}$ & 9$\times$10$^{-8}$ & 3$\times$10$^{-9}$ & 5.2$\pm$1.1$\times$10$^{-6}$ \\
A$_V$(int)\tablenotemark{j} & (mags)       & 35                 & 12                 & 25      & 7$\pm$0.5 \\
A$_V$(cir)\tablenotemark{k} & (mags)       & 0                  & 57                 & 26      & 96$\pm$8 \\
M$_{env}$\tablenotemark{l} & (M$_{\odot}$)                  & 8$\times$10$^{-9}$ & 1$\times$10$^{-2}$ & 4$\times$10$^{-1}$ & 1.8$\pm$0.2$\times$10$^{-2}$ \\
M$_{disk}$\tablenotemark{m} & (M$_{\odot}$)                 & 6$\times$10$^{-3}$ & 1$\times$10$^{-2}$ & 8$\times$10$^{-4}$ & 8.4$\pm$1.3$\times$10$^{-3}$ \\
\enddata

\tablenotetext{a}{Unconstrained best-fit values.}
\tablenotetext{b}{Best-fit values constraining M$_*\leq$2~M$_{\odot}$.}
\tablenotetext{c}{Best-fit values constraining M$_*\leq$2~M$_{\odot}$ and 
age between 10$^5$ and 10$^7$~yrs.}
\tablenotetext{d}{Values obtained from averaging the top 20 best-fit models with A$_V$(interstellar)$<$10~mags.}
\tablenotetext{e}{Radius of star.}
\tablenotetext{f}{Total luminosity = star+accretion luminosities.}
\tablenotetext{g}{Required disk inclination with respect to the line-of-sight.}
\tablenotetext{h}{Mass accretion rate from envelope to disk.}
\tablenotetext{i}{Mass accretion rate from disk to stellar surface.}
\tablenotetext{j}{Visual extinction not associated with circumstellar disk and envelope.}
\tablenotetext{k}{Visual extinction from circumstellar disk and envelope.}
\tablenotetext{l}{Envelope mass.}
\tablenotetext{m}{Disk mass.}

\end{deluxetable}

\clearpage

\begin{deluxetable}{lccccr}
\rotate
\tablecaption{Accretion Luminosity/Rates from Br$\gamma$ Emission Line Fluxes\label{accret}}
\tablewidth{0pt}
\tablehead{
\colhead{UT Date} & 
\colhead{Flux (observed)\tablenotemark{a}} &
\colhead{A$_V$\tablenotemark{b}} &
\colhead{Flux (dereddened)\tablenotemark{c}} &
\colhead{L$_{acc}$\tablenotemark{d}} &
\colhead{\.M$_{acc}$\tablenotemark{e}} \\
\colhead{} &
\colhead{(ergs~cm$^{-2}$~s$^{-1}$)} &
\colhead{(mags)} &
\colhead{(ergs~cm$^{-2}$~s$^{-1}$)} &
\colhead{(L$_{\odot}$)} &
\colhead{(M$_{\odot}$~yr$^{-1}$)}}
\startdata
2010 July 14\tablenotemark{f}     & 1.4(--14)\tablenotemark{g} & 12 & 4.8(--14) & 2.1 & 6.4(--7) \\
2010 July 18\tablenotemark{f}     & 2.9(--14) & 12 & 1.0(--13) & 5.2 & 1.6(--6) \\
2010 October 2   & 9.0(--14) &  2 & 1.1(--13) & 5.9 & 1.8(--6) \\
2010 November 26 & 2.3(--14) &  7 & 4.7(--14) & 2.0 & 6.3(--7) \\
\enddata
\tablenotetext{a}{Observed Br$\gamma$ emission line flux.}
\tablenotetext{b}{Visual extinction from JHK c-c diagram.}
\tablenotetext{c}{Dereddened Br$\gamma$ emission line flux.}
\tablenotetext{d}{Accretion luminosity.}
\tablenotetext{e}{Accretion rate.}
\tablenotetext{f}{From Covey et al. (2011).} 
\tablenotetext{g}{Number in parentheses is exponent e.g. 1.4(--14) 
means 1.4$\times$10$^{-14}$.}
\end{deluxetable}

\clearpage

\begin{deluxetable}{lccccr}
\rotate
\tablecaption{Mass-Loss Rates from [\ion{O}{1}] Emission Line Fluxes\label{massloss}}
\tablewidth{0pt}
\tablehead{
\colhead{UT Date} & 
\colhead{Flux} &
\colhead{A$_V$\tablenotemark{a}} &
\colhead{L$_{6300}$(observed)\tablenotemark{b}} &
\colhead{L$_{6300}$(dereddened)\tablenotemark{c}} &
\colhead{\.M$_{wind}$} \\
\colhead{} &
\colhead{(ergs~cm$^{-2}$~s$^{-1}$)} &
\colhead{(mags)} &
\colhead{(L$_{\odot}$)} &
\colhead{(L$_{\odot}$)} &
\colhead{(M$_{\odot}$~yr$^{-1}$)}}
\startdata
2010 July  8\tablenotemark{d}     & 5.5(--15)\tablenotemark{e} & 12 & 6.2(--5) & 4.4(--1) & 1.6(--6) \\
2010 July 19\tablenotemark{d}     & 4.0(--15) & 12 & 4.5(--5) & 3.3(--1) & 1.1(--6) \\
2010 September 5 & 1.4(--14) &  2 & 1.6(--4) & 7.1(--4) & 2.5(--9) \\
2010 November 25 & 4.9(--15) &  7 & 5.5(--5) & 1.0(--2) & 3.4(--8) \\
\enddata
\tablenotetext{a}{Visual extinction from JHK c-c diagram.}
\tablenotetext{b}{Observed emission line luminosity.}
\tablenotetext{c}{Dereddened emission line luminosity.}
\tablenotetext{d}{From Covey et al. (2011).} 
\tablenotetext{e}{Number in parentheses is exponent e.g. 5.5(--15) 
means 5.5$\times$10$^{-15}$.}
\end{deluxetable}

\clearpage

\clearpage

\clearpage

\clearpage

\clearpage

\clearpage

\clearpage

\clearpage

\clearpage

\clearpage

\clearpage

\clearpage

\begin{figure}[t] 
\begin{center}
\includegraphics*[angle=270,scale=1.3]{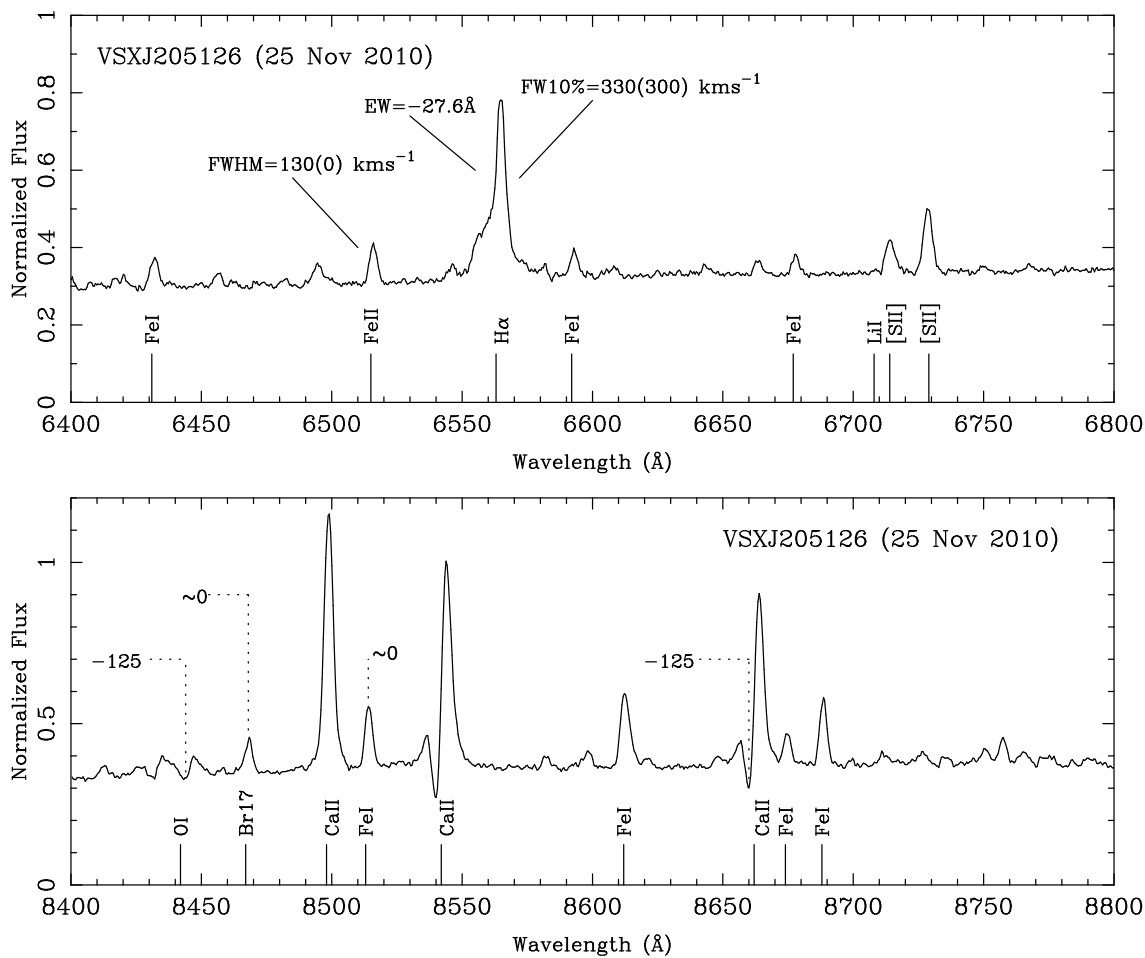} 
\caption{Top, the 6400--6800~\AA\ region of the spectra shown in
Figure~\ref{full2}.  H$\alpha$ is in emission but the associated 
blue-shifted absorption component seen on September 5 has disappeared. 
Bottom, the 8400--8800~\AA\ region of the spectra shown in 
Figure~\ref{full2}.  This region contains the far-red \ion{Ca}{2} 
triplet lines.  \vsx\ shows the far-red \ion{Ca}{2} triplet 
lines strongly in emission with blue-shifted absorption components visible
on two of the three lines.  The center of the absorption trough has decreased
from --190~km~s$^{-1}$ on September 5 to --125~km~s$^{-1}$ on UT 2010 
November 25.
\label{vhacaii2}}
\end{center}
\end{figure}
\clearpage

\begin{figure}[t] 
\begin{center}
\includegraphics*[angle=270,scale=0.65]{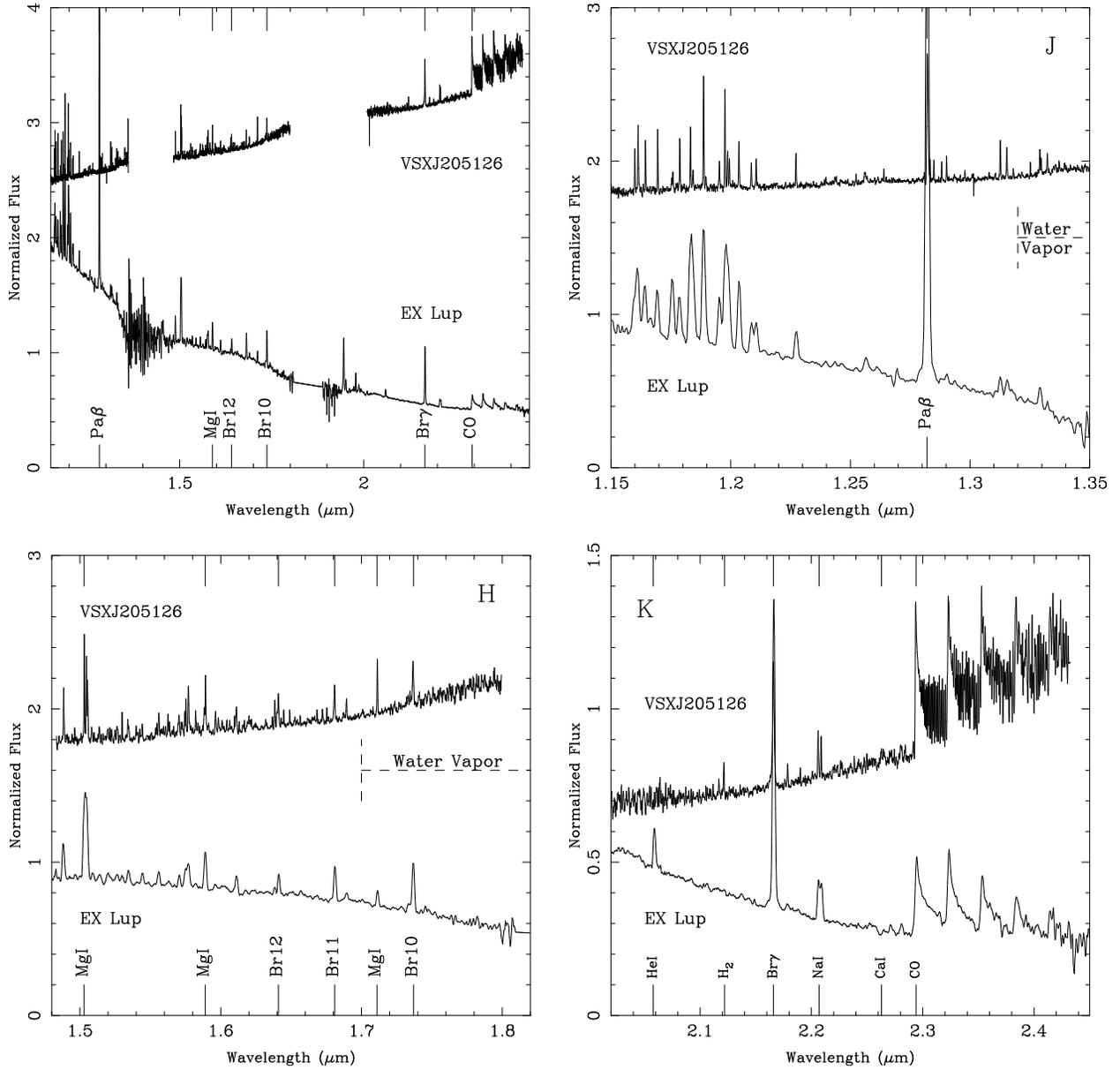} 
\caption{NIFS spectra of \vsx\ from UT 2010 October 2 and 3.  
Top-left is the full J, H, and K
  band spectra.  Top-right, bottom-left, and bottom-right show closer
  views of the J, H, and K bands, respectively.  For comparison, a 
lower resolution
  spectrum of EX~Lup (R$\sim$1200 taken during its 2008 outburst using the IRTF
  SpeX spectrograph) is also shown.  Note the very similar spectral
  structure in both sources including CO overtone emission bandheads, and 
  both Br$\gamma$ and Pa$\beta$ strongly in emission.  The forest of
  emission lines from 1.15--1.22~$\mu$m are predominantly from \ion{Fe}{1}.
  \label{compv-oct}}
\end{center}
\end{figure}
\clearpage

\begin{figure}[t] 
\begin{center}
\includegraphics*[angle=270,scale=0.9]{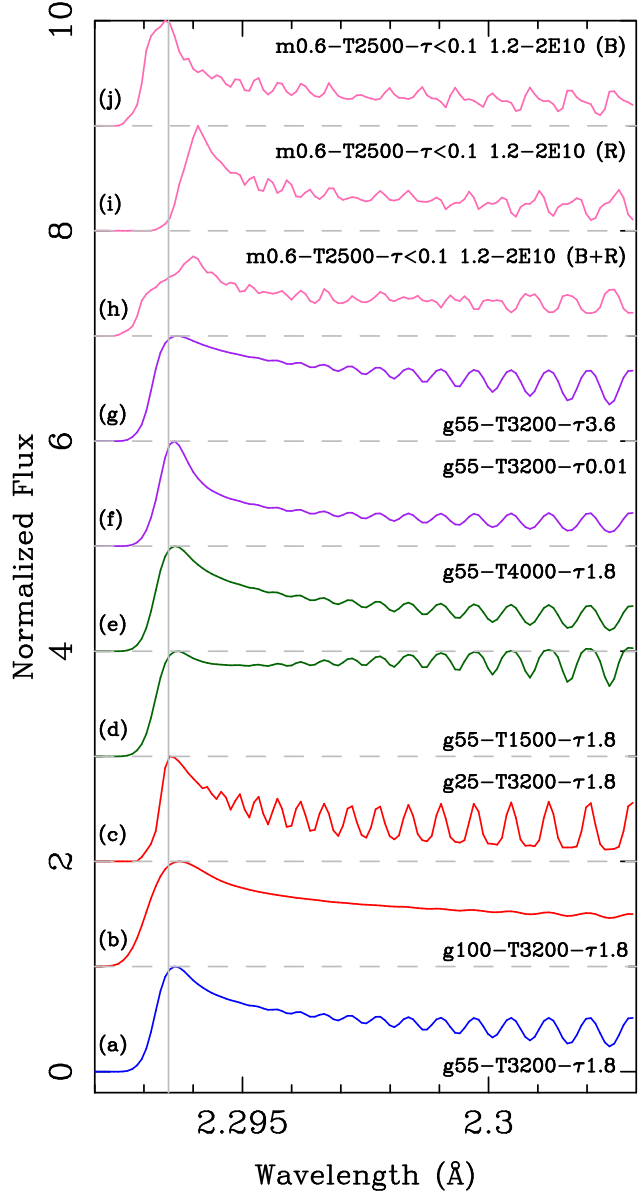} 
\caption{A selection of CO overtone bandhead emission models showing the effect
of changing the free parameters.   The parameters used to generate the model
are shown in each sub-plot.   Sub-plot (a, blue) shows the best-fit model to
the \vsx\ data shown in Figure~\ref{co1-oct} 
(e.g. Gaussian velocity broadening
{\it v$_g$}~=~55~km~s$^{-1}$, CO gas temperature T$_{CO}$~=~3200~K, and 
CO optical depth $\tau_{CO}$~=~1.8).   Sub-plots (b) and (c, both red) show
the effect of changing {\it v$_{g}$} to 100 and 25~km~s$^{-1}$, respectively. 
Sub-plots (d) and (e, both green) show the result of changing T$_{CO}$ to 1500
and 4000~K, respectively.  Sub-plots (f) and (g, purple) show the effect of
changing $\tau_{CO}$ to 0.01 and 3.6, respectively.  Sub-plot (h, pink) shows 
the changes that occur when a disk velocity profile is used 
rather than a Gaussian profile.  The parameters used here are
stellar mass M$_*$~=~0.6~M$_{\odot}$, T$_{CO}$=2500~K, $\tau_{CO}$~=~0.1, and
the inner ({\it r$_{in}$}) and outer ({\it r$_{out}$}) disk radius of 1.2 and
2.0$\times$10$^{10}$~m, respectively, and are the values used in the 
best-fit EX~Lup model shown in Figure~\ref{co1-oct}.  Plots (i) and (j, 
also pink)
show the effect of occulting the blue-shifted and red-shifted disk velocities,
respectively.   
\label{co2}}
\end{center}
\end{figure}
\clearpage

\begin{figure}[t] 
\begin{center}
\includegraphics*[angle=270,scale=0.7]{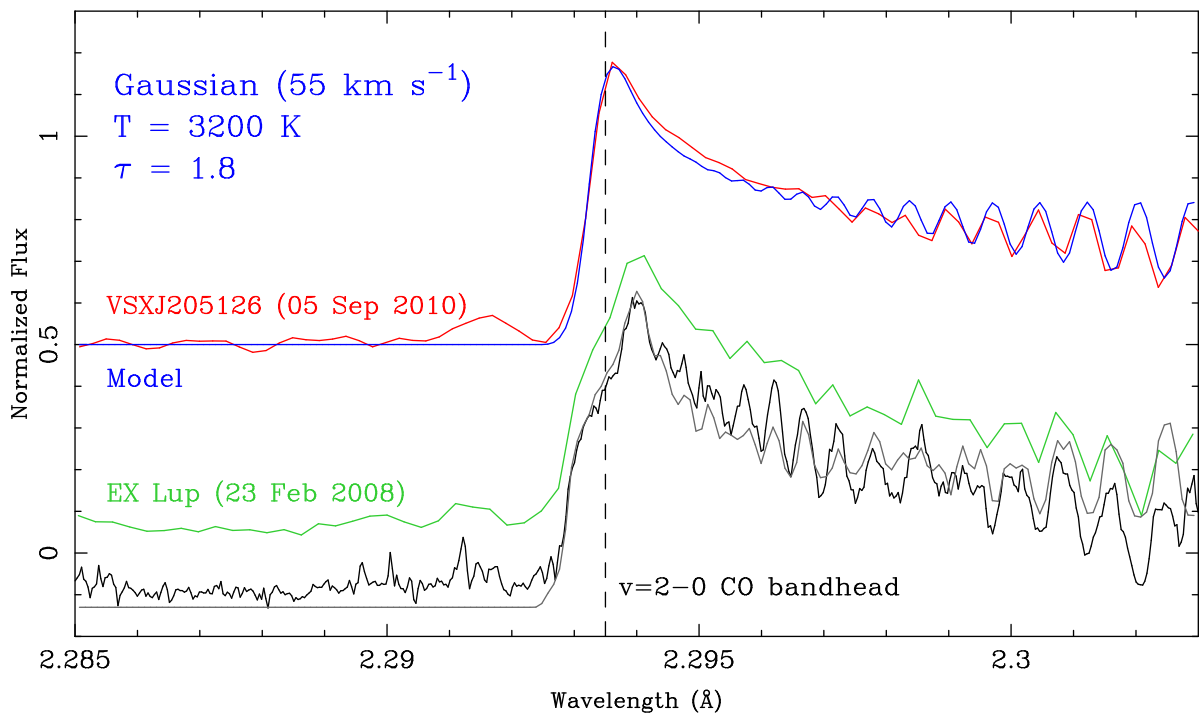} 
\caption{The v=2-0 CO overtone bandhead emission profile from UT October 2 
(red) overlaid by
the best-fit model (blue).  Also shown are observations and a model for EX~Lup
during its 2008 extreme outburst (from Aspin et al. 2010).  The EX~Lup data 
was taken at higher spectral resolution (black) but was binned to the 
resolution of the \vsx\ spectrum (green) to show that the structure 
present is still visible in the lower resolution version.  The best-fit model
to the EX~Lup data is also shown (grey).   The parameters used to produce 
the best-fit model to the \vsx\ spectrum are shown (top-left).   
\label{co1-oct}}
\end{center}
\end{figure}
\clearpage

\begin{figure}[t] 
\begin{center}
\includegraphics*[angle=270,scale=0.65]{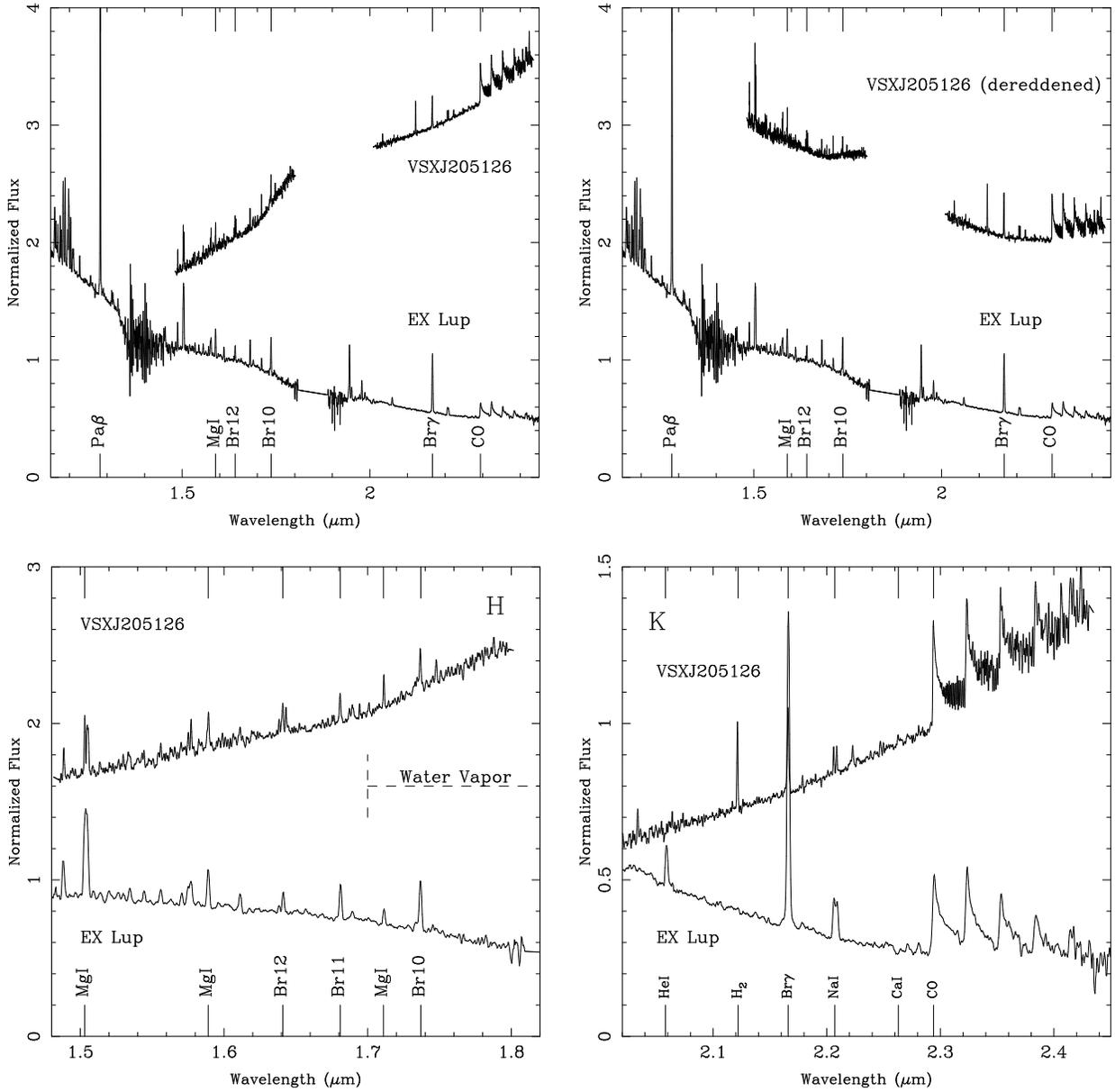} 
\caption{NIFS spectra of \vsx\ from UT November 26 and 27.  
Top-left is the full H and K band spectra (J was not obtained on this date).  
Bottom-left and bottom-right show closer views of the H, and K bands, 
respectively.  Again, the spectrum of EX~Lup is also shown. In the 
top-right panel we show a dereddened version of the full H to 
K-band spectrum of \vsx\ using an A$_V$=20~mags.  This demonstrates
that dereddening can closely match the continuum slope of EX~Lup which has
an intrinsic A$_V$ of 0--1~mags.
  \label{compv-nov}}
\end{center}
\end{figure}
\clearpage

\begin{figure}[t] 
\begin{center}
\includegraphics*[angle=270,scale=0.7]{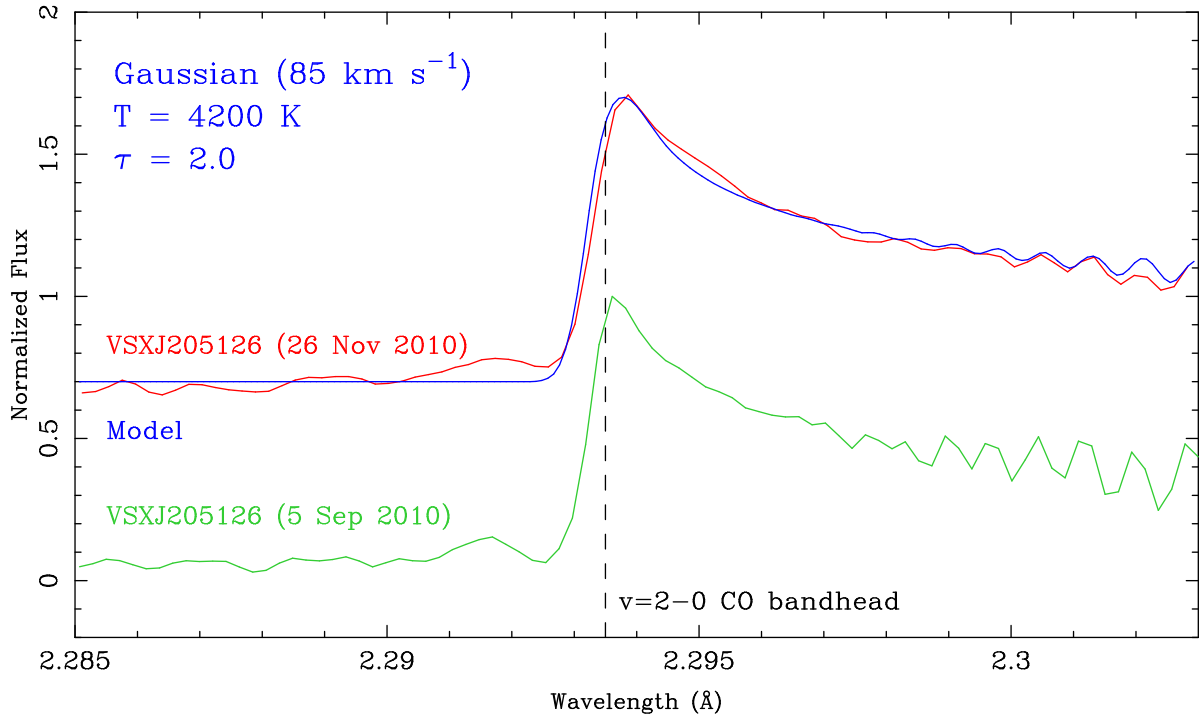} 
\caption{The v=2-0 CO overtone bandhead emission profile from UT November 26 
(red) overlaid by
the best-fit model (blue).  As in Figure~\ref{co1-oct}, observations 
(black and green) and 
best-fit model (grey) for EX~Lup are shown.  
The parameters used to produce 
the best-fit model to the \vsx\ spectrum are shown (top-left).   
\label{co1-nov}}
\end{center}
\end{figure}
\clearpage

\begin{figure}[t]
\begin{center}
\includegraphics*[angle=270,scale=0.65]{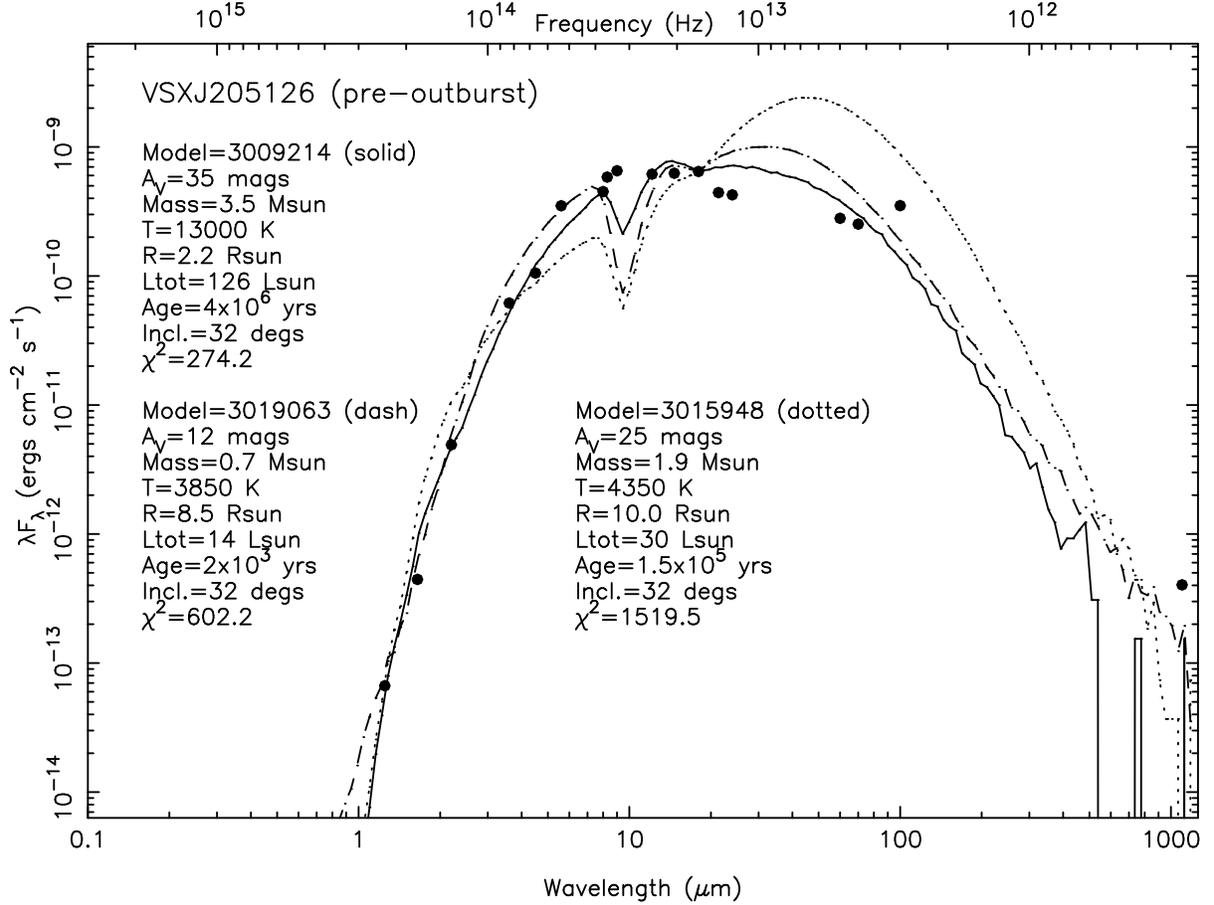} 
\caption{Pre-outburst spectral energy distribution (SED) for \vsx. 
The J, H, K' fluxes are from the UKIDSS database, the 8.28, 
9.0, 12.13, 14.65, 18.0, and 21.34~$\mu$m fluxes are from {\it MSX6C}, 
while the 60 and 100~$\mu$m fluxes are from {\it IRAS}.   Three models from
Robitaille et al. (2007) are overlaid on the data.  The solid line shows
the overall best fit.  The dashed line shows the best fit with M$_*$ 
constrained to be $<$2~M$_{\odot}$.  The dotted line shows the best fit
constrained with both M$_*<$2~M$_{\odot}$ and 10$^5<$t$_*<$10$^7$~yrs.
\label{seds}}
\end{center}
\end{figure}
\clearpage


\begin{thebibliography}{}

\bibitem[]{} Allen, L.E. et al. (2002) \apj, 566, 993

\bibitem[]{} Ambartsumian, V.A. 1971, Astrophysics, 7, 331

\bibitem[]{} Aspin, C., Reipurth, B., Herczeg, G.J. \& Capak, P., 2010,
  \apjl, 719, 50

\bibitem[]{} Aspin, C., Beck, T.L., \& Reipurth, B. 2008, AJ, 135, 423

\bibitem[]{} Aspin, C., \& Reipurth, B. 2009, AJ, 138, 1137

\bibitem[]{} Bally, J. \& Reipurth, B. 2003, AJ, 126, 893

\bibitem[]{} Briceno, C. et al. 2004, ApJ, 606, L123

\bibitem[]{} Covey, K. et al. 2011, in press

\bibitem[]{} Davies, R.L. et al. 1997, SPIE, 2871, 1099

\bibitem[]{} Dobashi, K. et al. 2005, PASJ, 57, S1

\bibitem[]{} Draine, B.T. \& McKee, C.F. 1993, ARA\&A, 31, 373

\bibitem[]{} Greene, T.P. \& Meyer, P. 1995, \apj, 450, 233

\bibitem[]{} Gutermuth, R.A., Megeath, S.T., Muzerolle, J., et al.
  2004, \apjs, 154, 374

\bibitem[]{} Gutermuth, R.A., Myers, P.C., Megeath, S.T., et al. 2008,
  \apj, 674, 336

\bibitem[]{} Hartmann, L., \& Kenyon, S.J. 1985, ApJ, 299, 462

\bibitem[]{} Hartmann, L. \& Kenyon, S.J. 1996, ARAA, 34, 207

\bibitem[]{} Herbig, G.H. 1966, Vistas in Astronomy, 8, 109

\bibitem[]{} Herbig, G.H. 1977, ApJ, 217, 693

\bibitem[]{} Herbig, G.H. 1989, in ESO Workshop on {\em Low Mass Star
Formation and Pre-Main Sequence Objects}, ed. B. Reipurth, p. 233

\bibitem[]{} Herbig, G.H. 1990, \apj, 360, 639

\bibitem[]{} Herbig, G.H. 2008, \aj, 135, 637

\bibitem[]{} Herbig, G.H. \& Harlan, E.A. 1971, IBVS, 543

\bibitem[]{} Herbig, G.H., Petrov, P.P., \& Duemmler, R. 2003, ApJ, 595, 384

\bibitem[]{} Hodapp, K.-W. et al. 2003, \pasp, 115, 1388

\bibitem[]{} Hook, I., Jorgensen, I., Allington-Smith, J.R., Davies,
R.L., Metcalfe, N., Murowinski, R.G., \& Crampton, D. 2004, PASP,
116, 425

\bibitem[]{} Itagaki, K. 2010, CBET, 2426

\bibitem[]{} Laugalys, V. et al. 2006, Baltic Astron., 15, 483

\bibitem[]{} Lawrence, A., et al. 2007, \mnras, 379, 1599

\bibitem[]{} Meinunger, L. 1969, {\it Mitt. \"uber Ver\"anderlichen
    Sterne}, 5, 47

\bibitem[]{} McGregor, P.J. et al. 2003, SPIE, 4841.1581

\bibitem[]{} Murakami, H., Baba, H., \& Barthel, P., et al. 2007,
  PASJ, 59, 369

\bibitem[]{} Osorio, M., D'Alessio, P., Muzerolle, J., Calvet, N., \&
  Hartmann, L. 2003, RevMexAA (Serie de Conferencias) 15, 142

\bibitem[]{} Petrov, P.P. \& Herbig, G.H. 1992, ApJ, 392, 209

\bibitem[]{} Petrov, P.P. \& Herbig, G.H. 2008, AJ, 136, 676

\bibitem[]{} Reipurth, B. \& Schneider, N. 2008, in {\em Handbook of
Star Forming Regions Vol. I}, ed. B. Reipurth, ASP, p. 36

\bibitem[]{} Reipurth, B. \& Aspin, C. 2004, ApJ, 606, L119

\bibitem[]{} Robitaille, T.P., Whitney, B.A., Indebetouw, R., Wood,
K., \& Denzmore, P. 2006, ApJS, 167, 256

\bibitem[]{} Robitaille, T.P., Whitney, B.A., Indebetouw, R., \& Wood,
K. 2007, ApJS, 169, 328

\bibitem[]{} Semkov, E. \& Peneva, S. 2010a, ATel, 2801

\bibitem[]{} Semkov, E. \& Peneva, S. 2010b, ATel, 2819

\bibitem[]{} Skrutskie, M.F. et al. 2006, AJ, 131, 1163

\bibitem[]{} Wachmann, A.A. 1954, Zs. f. Ap. 35, 74  

\bibitem[]{} Welin, G. 1971, A\&A, 12, 312

\bibitem[]{} Whitney, B.A., Kenyon, S.J., \& Gomez, M. 1997, \apj, 485, 703

\end{thebibliography}
\end{document}